\documentclass[conference]{IEEEtran}
\IEEEoverridecommandlockouts

\usepackage{cite}
\usepackage{amsmath,amssymb,amsfonts}
\usepackage{textcomp}
\usepackage{pifont}

\usepackage{graphicx}
\usepackage{xcolor}
\usepackage{algorithm}
\usepackage[noend]{algorithmic}
\usepackage{multirow}
\usepackage{multicol}
\usepackage{booktabs}
\usepackage{threeparttable}
\usepackage{comment}
\usepackage{bm}
\usepackage{enumerate}
\usepackage{diagbox}
\usepackage{flushend}
\usepackage{threeparttable}

\usepackage{amsthm}
\theoremstyle{plain}

\newtheorem{theorem}{Theorem}
\newtheorem{definition}{Definition}
\newtheorem{lemma}{Lemma}

\newtheorem{observation}{Observation}

\usepackage{subfigure}
\usepackage{setspace}
\usepackage{epstopdf}
\usepackage{hyperref}
\hypersetup{hidelinks}

\renewcommand{\algorithmiccomment}[1]{\bgroup\hfill$\triangleright$~#1\egroup}

\newcommand{\colorR}[1]{\textcolor{black}{#1}}

\def\BibTeX{{\rm B\kern-.05em{\sc i\kern-.025em b}\kern-.08em
    T\kern-.1667em\lower.7ex\hbox{E}\kern-.125emX}}
\begin{document}

\title{CARGO: Crypto-Assisted Differentially Private Triangle Counting without Trusted Servers}

\makeatletter
\newcommand{\linebreakand}{%
  \end{@IEEEauthorhalign}
  \hfill\mbox{}\par
  \mbox{}\hfill\begin{@IEEEauthorhalign}
}
\makeatother

\author{\IEEEauthorblockN{Shang Liu}
\IEEEauthorblockA{
% \textit{Dept. of Social Informatics} \\
\textit{Kyoto University}\\
 % Kyoto, Japan \\
shang@db.soc.i.kyoto-u.ac.jp}
\and
\IEEEauthorblockN{Yang Cao}
\IEEEauthorblockA{
% \textit{School of Information Science and Technology} \\
\textit{Hokkaido University}\\
% Sapporo, Japan \\
yang@ist.hokudai.ac.jp}
\and
\IEEEauthorblockN{Takao Murakami}
\IEEEauthorblockA{
% \textit{Dept. of Statistical Data Science} \\
\textit{Institute of Statistical Mathematics}\\
% Tokyo, Japan \\
tmura@ism.ac.jp}

\linebreakand 
\IEEEauthorblockN{Jinfei Liu}
\IEEEauthorblockA{
% \textit{Dept. of Computer Science and Technology} \\
\textit{Zhejiang University, ZJU-Hangzhou GSTIC}\\
% Hangzhou, China \\
jinfeiliu@zju.edu.cn}
\and
\IEEEauthorblockN{Masatoshi Yoshikawa}
\IEEEauthorblockA{
% \textit{Dept. of Data Science} \\
\textit{Osaka Seikei University}\\
% Osaka, Japan \\
yoshikawa-mas@osaka-seikei.ac.jp}
}

\maketitle

\begin{abstract}
Differentially private triangle counting in graphs is essential for analyzing connection patterns and calculating clustering coefficients while protecting sensitive individual information.
Previous works have relied on either central or local models to enforce differential privacy.
However, a significant utility gap exists between the central and local models of differentially private triangle counting, depending on whether or not a trusted server is needed.
In particular, the central model provides a high accuracy but necessitates a trusted server.
The local model does not require a trusted server but suffers from limited accuracy.
Our paper introduces a crypto-assisted differentially private triangle counting system, named CARGO, leveraging cryptographic building blocks to improve the effectiveness of differentially private triangle counting without assumption of trusted servers.
It achieves high utility similar to the central model but without the need for a trusted server like the local model.
CARGO consists of three main components. 
First, we introduce a similarity-based projection method that reduces the global sensitivity while preserving more triangles via triangle homogeneity.
Second, we present a triangle counting scheme based on the additive secret sharing that securely and accurately computes the triangles while protecting sensitive information.
Third, we design a distributed perturbation algorithm that perturbs the triangle count with minimal but sufficient noise.
We also provide a comprehensive theoretical and empirical analysis of our proposed methods.
Extensive experiments demonstrate that our CARGO significantly outperforms the local model in terms of utility and achieves high-utility triangle counting comparable to the central model.

\end{abstract}


\section{Introduction}
\label{sec:introduction}
Graph data analysis is gaining popularity in various fields such as social networks, transportation systems, and protein forecasting due to its widespread presence.
In graph analysis, triangle counting\cite{seshadhri2019scalable} is a crucial component for downstream tasks, including clustering coefficient \cite{newman2009random}, transitivity ratio \cite{schank2005approximating}, and structural similarity \cite{goldsmith1990assessing}.
However, triangle counting involves sensitive individual information that could be leaked through the results of the process \cite{tai2011privacy}. 
Differential privacy (DP) \cite{dwork2014algorithmic,li2016differential} has been widely used to provide formal privacy protection. 
Existing works on differentially private triangle counting \cite{ding2021differentially,karwa2011private,kasiviswanathan2013analyzing,imola2021locally, sun2019analyzing,imola2022communication} are mainly based on two DP models depending on the trust assumption of the server: \emph{central differential privacy} (CDP), which requires a trusted server, and \emph{local differential privacy} (LDP), which is preferable since it does not rely on a trusted server.

However, there is a significant utility gap between CDP-based \cite{ding2021differentially,karwa2011private,kasiviswanathan2013analyzing} and LDP-based \cite{ye2020lf,imola2021locally, sun2019analyzing,imola2022communication} of differentially private triangle counting, depending on whether or not there is a trusted server needed. 
In particular, CDP-based triangle counting models (as shown in Fig. \ref{fig:threemodels}(a)) need a trusted server to collect the whole graph before executing differentially private graph analysis.
For a graph with $n$ users and a privacy budget $\varepsilon$, 
the squared error of the central model is at most $O(\frac{d_{max}^2}{\varepsilon^2})$, 
where $d_{max}$ is the maximum degree in a graph.
In contrast, existing LDP-based triangle counting models (as shown in Fig. \ref{fig:threemodels}(c)) do not require a trusted server.
Instead, each user perturbs its sensitive information using an LDP mechanism and sends noisy data to the untrusted server.
The server then aggregates the data and releases a noisy triangle count.
However, LDP-based triangle counting models incur more error of $O(\frac{e^\varepsilon}{(e^\varepsilon-1)^2}(d_{max}^3n+\frac{e^\varepsilon}{\varepsilon^2}d_{max}^2n))$ (refer to Table 2 in \cite{imola2021locally} as the state-of-the-art LDP-based triangle counting protocol), which is much larger than that of CDP-based models, especially, when $n$ is large.

\begin{figure*}[t]
	\centering  
	\includegraphics[width=0.85\linewidth]{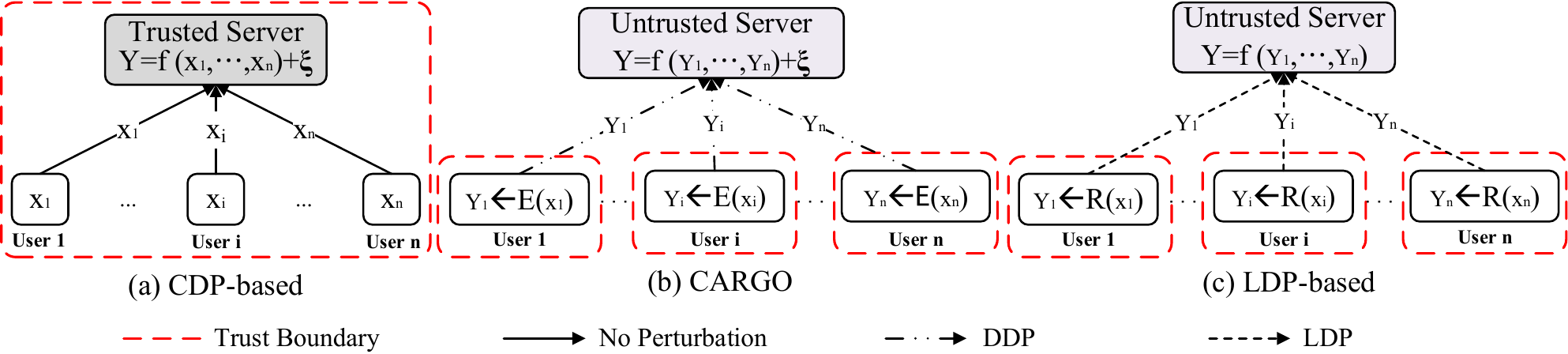}
 % \vspace{-0.3cm}
	\caption{A comparison among CDP-based model, LDP-based model and CARGO. (a) CDP-based model relies on a trusted server and achieves a high accuracy, e.g., $O(\frac{d_{max}^2}{\varepsilon^2})$ error; (c) LDP-based model removes a trusted server but introduces more error, i.e.,  $O(\frac{e^\varepsilon}{(e^\varepsilon-1)^2}(d_{max}^3n+\frac{e^\varepsilon}{\varepsilon^2}d_{max}^2n))$; (b) Our CARGO achieves a good utility like CDP-based model, i.e., $O(\frac{d_{max}^{\prime2}}{\varepsilon^2})$, but without a trusted server like LDP-based model, where $\varepsilon, n, d_{max}, d_{max}^\prime$ denote the privacy budget, number of users, true maximum degree, and noisy maximum degree, respectively.}
	\label{fig:threemodels}  
\end{figure*}

In this paper, we propose a \textbf{c}rypto-assisted differenti\textbf{a}lly p\textbf{r}ivate trian\textbf{g}le c\textbf{o}unting system
({CARGO}) that (1) achieves the high-utility triangle counting of the central model (2) without a trusted server like the local model.
Our goal is to calculate the triangles in a graph, where each node represents a user and each edge denotes the relationship between users, while protecting each user's neighboring information (i.e., edges).
Our system is inspired by recent studies  
\cite{he2017composing,roy2020crypt,roth2019honeycrisp,gu2021precad,stevens2022efficient, truex2019hybrid,SunL21,fu2022dp} that employ cryptographic techniques to bridge the utility gap between LDP and CDP models.
However, these systems are specifically designed to process tabular data \cite{he2017composing,roy2020crypt,roth2019honeycrisp} or gradients in federated learning \cite{gu2021precad,stevens2022efficient,truex2019hybrid} but not graph data.
Designing secure and private methods for counting triangles requires new principles due to the high sensitivity of triangle counting and the limited view of local users for a global graph.
As shown in Fig. \ref{fig:threemodels}(b), CARGO establishes a trust boundary for local data by leveraging cryptographic primitives and distributed differential privacy instead of injecting LDP noise, allowing for high-utility triangle counting comparable to that of CDP-based models (see more details in section \ref{sec:principle}).
We now elaborate on our key contributions:

\noindent\textbf{Similarity-based projection}.
We propose a novel \textit{local} projection method to reduce the  sensitivity of the triangle counting while preserving more triangles.
A significant obstacle in achieving differential privacy in counting triangles is the high sensitivity of triangle queries, leading to more noises needed for differentially private results.
The basic idea to address this in a central setting \cite{kasiviswanathan2013analyzing, day2016publishing} is to project (i.e., truncate) the original graph into a bounded graph.
The previous local graph projection method via randomly deleting edges \cite{imola2021locally} tries to reduce the sensitivity but results in much projection loss.
Our similarity-based projection method relies on the significant fact that node degrees of a triangle are pretty similar to each other \cite{durak2012degree}.
We prioritize deleting the edges with the least possibility of constructing triangles, which results in preserving more triangles.
It is worth noting that this simple yet efficient local projection algorithm can also be used to improve the utility of locally private triangle counting (details in Section \ref{subsec:projection}).

\noindent\textbf{ASS-based triangle counting}.
We introduce a novel triangle counting algorithm based on an additive secret sharing (ASS) technique~\cite{shamir1979share}.
Local users often face difficulties in calculating triangle counts due to their limited view of the global graph.
This limitation prevents them from seeing the third edges between others.
The state-of-the-art triangle counting method~\cite{imola2021locally} attempts to address this problem in untrusted settings by including an additional round of interaction.
However, there is still a significant gap in utility compared to the central model.
We propose a secure triangle counting method based on the ASS method with high accuracy.
A triangle exists in a graph if three edges of a triple exist simultaneously.
Namely, the multiplication of three bits in a matrix is equal to 1.
We introduce a protocol for multiplying three secret values, which allows us to securely and accurately compute the triangle counts while protecting sensitive neighboring information (details in Section \ref{subsec:local triangle count}).

\noindent\textbf{Distributed perturbation}.
We present a distributed perturbation method by combining additive secret sharing \cite{shamir1979share} and distributed noise generation \cite{shi2011privacy,acs2011have,goryczka2015comprehensive}.
The previous state-of-the-art crypto-assisted differential privacy (crypto-assisted DP) method \cite{roy2020crypt} randomizes the private value by adding two instances of Laplace noise, which leads to significant loss of utility.
Our distributed perturbation adds minimal but sufficient noise to the local user data. 
This partial noise is insufficient to provide an
LDP guarantee but the aggregated noise is enough to provide
a CDP protection.
Furthermore, secret sharing ensures that two untrusted servers only see encoded values beyond other information.
And the final aggregated noise can provide $\varepsilon$-Edge Distributed Differential Privacy (DDP) guarantee (details in Section \ref{subsec:randomization via DDP}).

\noindent\textbf{Comprehensive theoretical and empirical analysis}.
We provide a comprehensive theoretical analysis of our proposed protocols, including utility, privacy, and time complexity.
In particular, we provide the upper-bounds on the estimation error for triangle counting and find that CARGO can significantly reduce the estimation error of local models.
Additionally, we prove that our proposed CARGO satisfies $\varepsilon$-Edge DDP (details in Section \ref{sec:theoretical analysis}).
Finally, several experiments have been conducted to demonstrate that our CARGO achieves high-utility triangle counts comparable to central models, and significantly outperforms local models by at least an order of 5 in utility (details in Section \ref{sec:experiment}). 

\section{Preliminaries}
\label{sec:preliminary}

\subsection{Problem Statement}
\label{subsec:problem statement}
\subsubsection{Graphs and Triangle Counting}
In our work, we consider an undirected graph with no additional attributes on nodes or edges, which can also be represented as $G=(V, E)$, where $V=\{v_1,...,v_n\}$ is the set of nodes, and $E\subseteq V \times V$ is the set of edges. 
Each local user $v_i$ owns one adjacent bit vector $A_i=\{a_{i1},...,a_{in}\}$ that records the neighboring information, where $a_{ij}=1, j\in[n]$ if and only if edge $\langle v_i,v_j \rangle\in E$.
The adjacent bit vectors of all local users compose a symmetric adjacency matrix $A=\{A_1, A_2,..., A_n\}$.
A triangle in a graph $G$ consists of three nodes where each node connects to the other two nodes.
Table \ref{tab:notations} summarizes the major notations used in this paper.

\begin{table}[t]
\small
	\caption{Summary of Notations.}
	\label{tab:notations}
	\centering
	\setlength{\tabcolsep}{5mm}{
		\begin{tabular}{l|l}
			\hline
			Notation & Definition \\
			\hline
	$G=(V,E)$ & Graph with nodes $V$ and edges $E$\\
	$n$ & Number of users \\
        $v_i$ & $i$-th node in $V$ \\
        $d_i$ & Node degree of $v_i$ \\
        $ A$ & Adjacent matrix\\
        $ A_i$ & Adjacent bit vector of $v_i$ \\
        $ D$ & True degree set \\
        $ D^\prime$ & Noisy degree set\\
        $d_{max}$ & True maximum degree \\
        $d_{max}^\prime$ & Noisy maximum degree\\
        $\triangle$ & Sensitivity of triangle counting\\
        $ T$ & True number of triangles \\
        $T^\prime$ & Noisy number of triangles\\
        %$\langle x \rangle$ & Secret sharing of $x$\\
	\hline
	\end{tabular}}
\end{table}

\subsubsection{Trust Assumptions}
Our system includes $n$ users and two servers, as illustrated in Fig. \ref{fig:framework}.
We assume that two servers are \emph{semi-honest} and \emph{non-colluding}.
This is a common assumption in cryptographic systems, such as \cite{roy2020crypt,patra2021aby2,rathee2020cryptflow2, gu2021precad}, and can be enforced via strict legal bindings.
Semi-honest implies that they follow the protocol instructions honestly but may be curious about additional information. 
Non-colluding means that they do not disclose any information to each other beyond what is allowed by the defined protocol. 
Furthermore, we assume that there are no corrupt users, and each user has a private channel with each server to share sensitive information confidentially. 
We also assume that any parties beyond the system, such as servers, analysts, or other individuals, are adversaries who are computationally constrained. 

\subsubsection{Utility metrics}
We use two common utility metrics to evaluate our methods, including $l_2$ loss (e.g., squared error) like \cite{murakami2019utility,wang2017locally}, and relative error as with \cite{chen2012differentially,bindschaedler2016synthesizing}.
To be specific, let $T^\prime$ be a private estimation of the true triangles $T$.
The $l_2$ loss function maps the true number of triangles $T$ and the private estimation $T^\prime$ to the $l_2$ loss, which can be denoted by: $l_2(T,T^\prime)=(T-T^\prime)^2$.
When $T$ is large, the $l_2$ loss may also be large.
Thus, we also compute the relative error in our experiments.
The relative error is defined as: $re(T,T^\prime)=\frac{|T-T^\prime|}{T}$, where $T \neq$ 0.

\subsection{Differential Privacy on Graphs}
\label{subsec:DP_Graphs}

Differential privacy (DP)  \cite{dwork2014algorithmic,li2016differential} has become a standard for privacy protection, which can formalized in Definition \ref{def:CDP}.
Based on different trusted assumptions, DP can be divided into two types: \emph{central differential privacy} (CDP) and \emph{local differential privacy} (LDP).

\begin{definition}[Differential Privacy \cite{dwork2014algorithmic}]
	\label{def:CDP}
Let $n$ be the number of users.
Let $\varepsilon > 0$ be the privacy budget.
Let $\mathcal{X}$ be the set of input data for each user.
A randomized algorithm $\mathcal{M}$ with domain $\mathcal{X}^n$ satisfies $\varepsilon$-DP, iff for any neighboring databases $D, D^\prime \in \mathcal{X}^n$ that differ in a single datum and any subset $S \subseteq Range(\mathcal{M})$, 
	\begin{center}
		$Pr[\mathcal{M}(D) \in S] \leq e^{\epsilon} Pr[\mathcal{M}(D^{\prime}) \in S]$
	\end{center}  
\end{definition}

In this work, we use the global sensitivity \cite{dwork2014algorithmic} to achieve the DP. 
The global sensitivity considers the maximum difference between query results on two neighboring databases. 

\textbf{Edge DP}.
As a graph consists of nodes and edges, there are two definitions when DP is applied to either of them: \emph{edge differential privacy} (Edge DP) \cite{hay2009accurate} and \emph{node differential privacy} (Node DP) \cite{hay2009accurate}.
Edge DP guarantees the output of a randomized mechanism does not reveal whether any friendship information (i.e., edge) exists in a graph $G$;
whereas Node DP hides the existence of one user (e.g., node) along with her adjacent edges.
Node DP provides a stronger privacy guarantee since it protects not only edge information but also node information.
But Node DP brings much more error than Edge DP.
Considering our privacy goal (protecting the neighboring information) and higher accuracy requirement, we choose to use Edge DP in our work like \cite{ding2021differentially,imola2021locally,ye2020lf, sun2019analyzing,imola2022communication}.

Definition \ref{def:edgeDP} and Definition \ref{def:edgeLDP} give the formal definition of Edge CDP and Edge LDP respectively.
Edge LDP assumes that two edges $\langle v_i,v_j\rangle$ and $\langle v_j,v_i\rangle$ between user $v_i$ and $v_j$ are different secrets~\cite{imola2021locally}.

\begin{definition}[Edge CDP \cite{raskhodnikova2016differentially}]
	\label{def:edgeDP}
Let $\varepsilon > 0$ be the privacy budget.
A randomized algorithm $\mathcal{M}$ with domain $\mathcal{G}$ satisfies $\varepsilon$-Edge CDP, iff for any two neighboring graphs $G, G^\prime \in \mathcal{G}$ that differ in one edge and any subset $S \subseteq Range(\mathcal{M})$, 
	\begin{center}
		$Pr[\mathcal{M}(G) \in S] \leq e^{\epsilon} Pr[\mathcal{M}(G^\prime) \in S]$
	\end{center} 
\end{definition}

\begin{definition}[Edge LDP \cite{qin2017generating}]
	\label{def:edgeLDP}
Let $\varepsilon > 0$ be the privacy budget.
For any $i\in[n]$, let $\mathcal{M}_i$ be a randomized algorithm of user $v_i$. 
$\mathcal{M}_i$ satisfies $\varepsilon$-Edge LDP, iff for any two neighboring adjacent bit vectors $A_i$ and $A_i^\prime$ that differ in one edge and any subset $S \subseteq Range(\mathcal{M}_i)$, 
	\begin{center}
		$Pr[\mathcal{M}_i(A_i) \in S] \leq e^{\epsilon} Pr[\mathcal{M}_i(A_i^\prime) \in S]$
	\end{center} 
\end{definition}

\textbf{Edge DDP}.
Combining DDP \cite{shi2011privacy,acs2011have,goryczka2015comprehensive} with standard Edge CDP (Definition \ref{def:edgeDP}), we introduce an Edge Distributed Differential Privacy (Edge DDP) for protecting triangle counting in CARGO.
The formal definition is as follows:

\begin{definition}[Edge DDP]
\label{def:Edge DDP}
Let $\varepsilon > 0$ be the privacy budget.
Let $r_i$ be a distributed noise generated by user $v_i$.
    A randomized algorithm $M$ with randomness over the joint distribution of $\bm{r}$:=$(r_1,...,r_n)$ satisfies $\varepsilon$-Edge DDP, iff for any neighboring graphs $G$ and $G^\prime$ that differ in one edge, for any output $y\in range(\mathcal{M})$,
    \begin{center}
        $Pr[\mathcal{M}(G)=y]\leq e^\varepsilon Pr[\mathcal{M}(G^\prime)=y]$
    \end{center}
\end{definition}

Note that both $\varepsilon$-Edge LDP (Definition~\ref{def:edgeLDP}) and $\varepsilon$-Edge DDP (Definition~\ref{def:Edge DDP}) protect one edge with privacy budget $\varepsilon$. 
The difference is that Edge LDP is for the local model, whereas Edge DDP is for the crypto-assisted DP model. 
We use Edge LDP to prove Edge DDP for the entire process of our CARGO.

\subsection{Additive Secret Sharing}
\label{subsec:crypto_preliminary}
In two-party additive secret sharing (ASS) \cite{shamir1979share}, a private value is split into two secret shares that can be used to construct the true value, like \cite{mohassel2017secureml,riazi2018chameleon, zheng2023secure}.
Each value is represented as an $l$-bit integer in the ring $\mathbb{Z}_{2^l}$, and cannot reveal any information about the private value. 
In this paper, we denote a secret share of $x$ by $\langle x \rangle$.
To additively share a secret value $x$, a random number $r \in \mathbb{Z}_{2^l}$ is generated.
Then the shares for parties $S_1$ and $S_2$ can be represented as $\langle x \rangle_1 =r$ mod $2^l$, $\langle x \rangle_2 = (x-r)$ mod $2^l$, respectively, where $\langle x \rangle=\langle x \rangle_1 + \langle x \rangle_2$. 
The basic operations naturally supported in the ASS domain include addition and multiplication.
Given two shared values $\langle x \rangle$ and $\langle y \rangle$, each party $S_{i\in \{1,2\}}$ receives $\langle x \rangle_i$ and $\langle y \rangle_i$.
Each party $S_i$ locally computes $\langle u \rangle_i$ = $\langle x \rangle_i$ + $\langle y \rangle_i$.
Then, the addition can be securely computed by aggregating $\langle u \rangle_1$ and $\langle u \rangle_2$, i.e., $u= \langle u \rangle_1 + \langle u \rangle_2  = \langle x \rangle_1$ + $\langle x \rangle_2 + \langle y \rangle_1 + \langle y \rangle_2 = x + y$.
The multiplication of two shared values may be complex.
It needs one-round communication of the Beaver triple, which can be prepared offline.

\begin{figure}[t]
	\centering  
	\includegraphics[width=\linewidth]{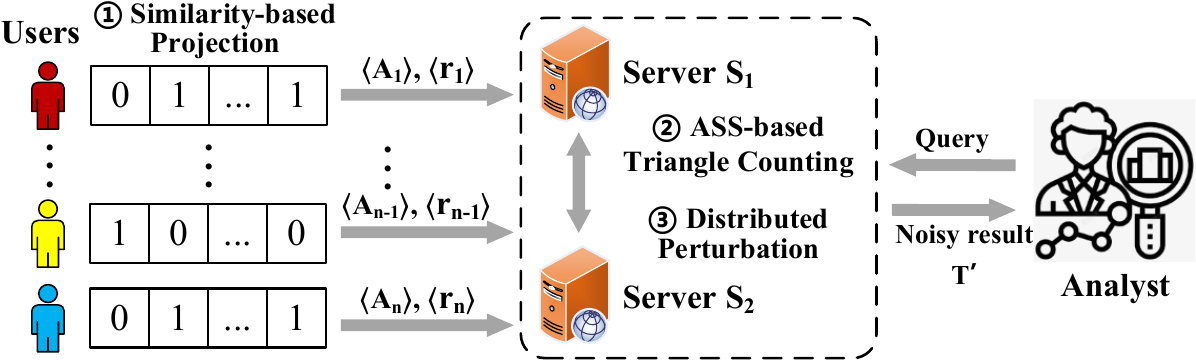}
	\caption{CARGO system.}
	\label{fig:framework}  
 % \vspace{-0.3cm}
\end{figure}

\section{Cargo System}
\label{sec:CARGO}

\subsection{Design principle}
\label{sec:principle}
Our main idea is to let the users and two servers \textit{collaboratively} compute secret shares of the true triangle counts securely and add distributed differentially private noise without requiring any trusted servers.
The previous crypto-assisted differentially private data analysis protocols
\cite{he2017composing,roy2020crypt,roth2019honeycrisp,gu2021precad,stevens2022efficient, truex2019hybrid} are designed for the general tabular data.
The high sensitivity of triangle counting and the limited view of local users for a global graph make these protocols less effective for private graph data analysis.
Furthermore, Crypt$\varepsilon$ \cite{roy2020crypt}, a state-of-the-art crypto-assisted DP protocol, employs two non-colluding and untrusted servers to independently add Laplace noise twice, resulting in twice the utility loss compared to CDP models. 
To this end, we first design a novel local projection based on triangle homogeneity to reduce the high sensitivity from $O(n)$ to $O(d_{max}^\prime)$ while preserving more triangles as much as possible.
Then, we propose a secure protocol for counting triangles using additive secret sharing techniques, which enables users and two servers to collaboratively calculate the secret shares of true triangle counts $T$, i.e., $\langle T \rangle_1$ and  $\langle T \rangle_2$.
We then introduce a distributed perturbation algorithm that adds minimal yet sufficient noise to triangles for privacy preservation. To safeguard the privacy of the partial noise, users do not directly transmit it to the server. Each user encodes the noise using additive secret sharing and then distributes it to the servers. The servers integrate this encoded noise into their secret shares of the triangle counts. By aggregating these shares, the servers can compute the differentially private triangle counts accurately.

\subsection{Framework}
Fig. \ref{fig:framework} shows CARGO's system architecture.
CARGO involves two kinds of entities: local users and two non-colluding servers.
The local user $v_i$, $i\in[n]$ ($n$ = number of users in a graph), owns the sensitive friendship information which is represented as an adjacent bit vector.
At the beginning, local users interact with one of the servers for the private estimation of maximum degree $d_{max}^\prime$.
The local user then projects the original adjacent bit vector into $d_{max}^\prime$-bounded adjacent bit vector (step \ding{172}).
Next, the local user secretly shares her adjacent bit vector to two servers and each server computes the secret share of the true triangle counts (step~\ding{173}).
Subsequently, the local user generates a distributed
noise and secretly shares it with two servers.
Two servers sum up their own shares of the noise, and add aggregated noise into the share of the triangle counts, respectively.
In other words, each server obtains the secret share of the noisy triangle counts personally.
The final aggregation of two shares is equal to the noisy triangle counts of the entire graph, which satisfies $\varepsilon$-Edge Distributed DP (step~\ding{174}).

\textbf{Main Steps}. 
Algorithm \ref{alg:overall protocol} presents the overall protocol of CARGO system, which consists of three main steps:

\begin{algorithm}[t]
\small
	\caption{Overall protocol of CARGO system} 
	\label{alg:overall protocol} 
	\begin{algorithmic}[1] 
 
		\REQUIRE  %Input 
       \begin{tabular}[t]{l}
         $G$ represented as adjacent lists $ A=\{ A_1,..., A_n\}$,\\ 
         True degree set $ D=\{d_1,...,d_n\}$,\\
         Privacy budget $\varepsilon=\varepsilon_1+\varepsilon_2$\\
      \end{tabular} \\
      
		\ENSURE  %Output 
		Noisy triangle count $T^\prime$

            \STATE Initialize: $ T=\varnothing$ \\
            \textbf{Step 1: Similarity-based Projection}\\
            \STATE 
            $(D^\prime,d_{max}^\prime) \leftarrow \mathsf{Max}(D,\varepsilon_1)$ \COMMENT{Algorithm \ref{alg:maximum degree}} \\

            \STATE $\hat{A} \leftarrow \mathsf{Project}(A, D, D^\prime, d_{max}^\prime)$
            \COMMENT{Algorithm \ref{alg:local projection}} \\

            \textbf{Step 2: ASS-based Triangle Counting}\\
            \STATE $\langle T \rangle \leftarrow \mathsf{Count}(\hat{A})$
            \COMMENT{Algorithm \ref{alg:local triangle counting}}\\

            \textbf{Step 3: Distributed Perturbation}
            \STATE $T^\prime \leftarrow \mathsf{Perturb}(\langle T \rangle,d_{max}^\prime,\varepsilon_2)$
            \COMMENT{Algorithm \ref{alg:randomization with distributed DP}} \\
            \RETURN $T^\prime$
	\end{algorithmic} 
\end{algorithm}

Step 1: \emph{Similarity-based Projection}.
Graph projection is the key technique to reduce the global sensitivity from $O(n)$ to $O(\theta)$, where $\theta$ is the projection parameter.
Here, we set $\theta$ as the maximum degree $d_{max}$ to avoid removing edges from an adjacent bit vector (e.g., to avoid the loss of utility).
However, there is no prior knowledge about the maximum degree.
CARGO recalls a $\mathsf{Max}(.)$ function (Algorithm \ref{alg:maximum degree}) to privately compute a noisy maximum degree $d_{max}^\prime$ that is approximately equal to $d_{max}$, as shown in Table \ref{tab:d_max_prime}.
Next, each user transforms the original adjacent bit vector into a $d_{max}^\prime$-bounded adjacent bit vector using a $\mathsf{Project}()$ function (Algorithm \ref{alg:local projection}).
Previous local projection method via randomly deleting edges \cite{imola2021locally} results in much projection loss.
We propose a similarity-based projection method by leveraging the triangle homogeneity (Observation~\ref{lem:degree correlations of a triangle}).
During the projection, the user deletes the edges with the least possibility of constructing triangles, and thus more triangles are preserved (see more details in Section \ref{subsec:projection}).

Step 2: \emph{ASS-based Triangle Counting}.
Next, the true triangle count can be computed based on the projected adjacent bit vector.
The main challenge in triangle counting is that each user has a limited view of a global graph.
In other words, local users cannot see the third edge between others. 
The previous state-of-the-art two-round triangle counting method \cite{imola2021locally} in untrusted settings leads to more errors.
We propose an ASS-based triangle counting method for securely and accurately calculating the triangle count.
Each user encodes each bit in its adjacent bit vector via additive secret sharing and sends it to two sever.
Each server obtains the secret share of the triangle count and knows nothing about the true result. 
CARGO recalls a $\mathsf{Count}()$ function (Algorithm \ref{alg:local triangle counting}) to calculate the number of triangle counts securely and accurately via Additive Secret Sharing (ASS) 
(refer to Section~\ref{subsec:local triangle count} for more details).

Step 3: \emph{Distributed Perturbation}.
After accurately calculating the shares of triangle counts, CARGO employs a $\mathsf{Perturb}()$ function (Algorithm \ref{alg:randomization with distributed DP}) to privately estimate the triangle count of a graph. 
The previous state-of-the-art crypto-assisted DP method \cite{roy2020crypt} guarantees differential privacy by incorporating two instances of Laplace noise, which results in a significant loss of utility.
We propose a distributed perturbation method by combining the additive secret sharing and distributed noise.
Each user first generates a minimal but sufficient noise.
Since such a small amount of noise is unable to provide enough privacy guarantees compared with LDP, we employ ASS to let each user split the generated noise into two secrets and share them with two servers, respectively.
The servers, in turn, are unable to decipher any information about the noise independently. By merging the shared noise with the shared triangle count, each server acquires a secret share of the noisy triangle count. The final computation of the noisy triangle count, safeguarded under $\varepsilon$-Edge Distributed DP by aggregating two shared noisy results (details in Section \ref{subsec:randomization via DDP}).

\textbf{Extension to Node DP}.
CARGO can be extended to Node DP by revising Algorithm \ref{alg:maximum degree} and Algorithm \ref{alg:randomization with distributed DP}.
The main change is from the sensitivity updates.
To be specific, given the number of nodes $n$, in Algorithm \ref{alg:maximum degree}, any change of one node will influence the other $(n-1)$ node degrees in the worst case.
Thus, the sensitivity of $\mathsf{Max}$ is $O(n)$ when we use Node DP.
Similarly, the sensitivity of $\mathsf{Perturb}$ in Algorithm \ref{alg:randomization with distributed DP} becomes $O\binom{d_{max}^\prime}{2}$.
Although our algorithm $\mathsf{Project}$ can reduce the high sensitivity from $O\binom{n}{2}$ to $O\binom{d_{max}^\prime}{2}$, there are still much more utility loss than Edge DP.
Therefore, how to reduce the high sensitivity of Node DP while preserving more triangles is the focus of future work.

\begin{algorithm}[t]
\small
	\caption{$\mathsf{Max}$: private estimation of $d_{max}$} 
	\label{alg:maximum degree} 
	\begin{algorithmic}[1] 
 
		\REQUIRE  %Input 
       \begin{tabular}[t]{l}
         True degree set $ D=\{d_1,...,d_n\}$,\\
         Privacy budget $\varepsilon_1$ \\
      \end{tabular} \\
      
		\ENSURE  %Output 
		Noisy maximum degree $d_{max}^\prime$

            \STATE Initialize: $ D^\prime=\varnothing$
            \FOR{$i=1$ \textbf{to} $n$}       
            \STATE $d_i^\prime \leftarrow d_i + \mathsf{Lap}(\frac{1}{\varepsilon_1})$ 
            \STATE $ D^\prime \leftarrow  D^\prime \cup \{d_i^\prime\}$
            \STATE Send $d_i^\prime$ to untrusted server
            \ENDFOR
            \STATE Server: $d_{max}^\prime \leftarrow \mathsf{max} (d_1^\prime,...,d_n^\prime)$
            \RETURN ($ D^\prime, d_{max}^\prime$)
	\end{algorithmic} 
\end{algorithm}

\subsection{Similarity-based Projection}
\label{subsec:projection}

\subsubsection{Private Estimation of the Maximum Degree}
In this work, we assign the maximum degree $d_{max}$ as the projection parameter $\theta$ just like \cite{imola2021locally,day2016publishing}, primarily for two main reasons.
On the one hand, as shown in Table \ref{tab:datasets}, $d_{max}$ is much smaller than the number $n$ of users in real-world graphs, which can significantly reduce the sensitivity.
On the other hand, it avoids deleting neighboring friends from an adjacent list ideally; i.e., it avoids the utility loss during the projection.
In the untrusted scenario, however, each user knows no about $d_{max}$ since it has a limited view of the global graph. 
To handle this, local users privately compute $d_{max}$ with the leverage of the server's global view.
Specifically, as shown in Algorithm \ref{alg:maximum degree}, each user $v_i$ first adds $\mathsf{Lap(\frac{1}{\varepsilon_1})}$ to her node degree $d_i$.
Here, we use Edge LDP (Definition \ref{def:edgeLDP}), and the sensitivity is one since two edges $\langle v_i,v_j\rangle$ and $\langle v_j,v_i\rangle$ between user $v_i$ and $v_j$ are different secrets~\cite{imola2021locally}, and any change of one edge will influence one node degree.
Then each user sends noisy degree $d_i^\prime$ to the untrusted server.
In CARGO, one of two parties $S_1$ and $S_2$ can be used for computing $d_{max}^\prime$.
Finally, the server computes the max value of the noisy degree sequence $\{d_1^\prime,...,d_n^\prime\}$ as $d_{max}^\prime$, and sends $d_{max}^\prime$ back to local users.
We denote this algorithm by $\mathsf{Max}$.

\subsubsection{Local Graph Projection}

\begin{figure}[t]
	\centering  
	\includegraphics[width=0.7\linewidth]{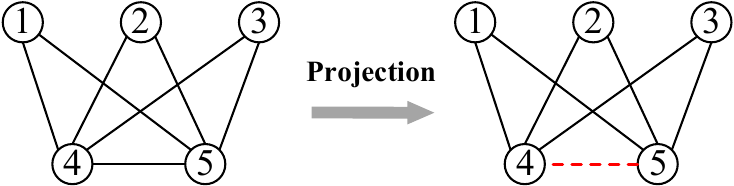}
	\caption{Limitation of random deletion. Assume that $d_{max}^\prime$ is equal to 3, if user $v_4$ (or $v_5$) projects the adjacent list by deleting the edge $\langle v_4,v_5 \rangle$, all triangles in a graph will be removed.
	\label{fig:motivation_projection}}
     %\vspace{-0.2cm}
\end{figure}

\begin{figure}[t]
	\centering  
	\includegraphics[width=0.9\linewidth]{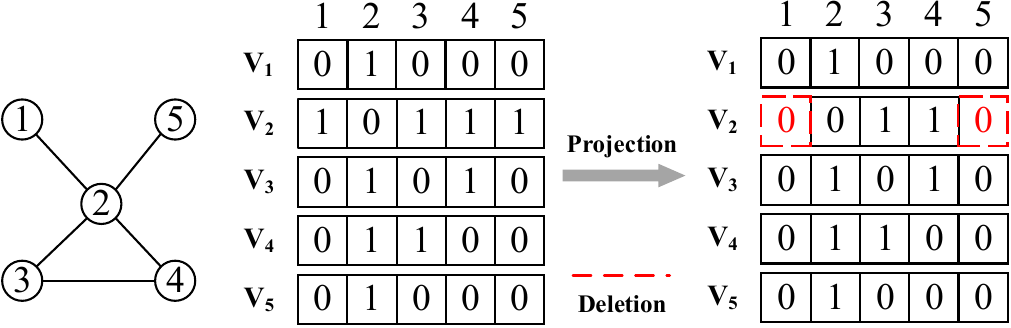}
	\caption{Local graph projection. Assume that $d_{max}^\prime$ is equal to 2, user $v_2$ projects the adjacent bit vector by deleting the edge $\langle v_2,v_1 \rangle$ and edge $\langle v_2,v_5 \rangle$.}
	\label{fig:local graph projection}  
\end{figure}

After obtaining the estimation of the maximum degree $d_{max}^\prime$, each user $v_i$ transforms the original adjacent bit vector $A_i$ into a $d_{max}^\prime$-bounded adjacent bit vector $\hat{A_i}$ via the graph projection.
Although there have been some works involving graph projection methods \cite{liucrypto,imola2021locally,day2016publishing,ding2021differentially}, they do not perform in the triangle counting within untrusted settings very well.
For instance, the most related work is that Imola et al. \cite{imola2021locally} implements graph projection via randomly deleting edges in untrusted settings.
However, this random projection possibly deletes key edges involved in many triangles.
For example, in Fig. \ref{fig:motivation_projection}, if user $v_4$ randomly deletes the edge $\langle v_4,v_5 \rangle$ to bound the adjacent list, all triangles in a graph will disappear.

We propose a similarity-based projection for reducing the global sensitivity in untrusted scenarios.
The main target of our methods is that the high global sensitivity can be reduced while more triangles can be preserved after the projection.
As illustrated in Fig. \ref{fig:local graph projection}, if $d_i > d_{max}^\prime$, user $v_i$ will delete $(d_i-d_{max}^\prime)$ friends from her adjacent bit vector.
The candidates to be deleted are selected based on significant knowledge in the following observation:

\begin{algorithm}[t]
\small
	\caption{$\mathsf{Project}$: Similarity-based Projection} 
	\label{alg:local projection} 
	\begin{algorithmic}[1] 
 
		\REQUIRE  %Input 
       \begin{tabular}[t]{l}
         Adjacent matrix $A$ of a graph\\
         True degree set $ D=\{d_1,...,d_n\}$ \\
         Noisy degree set $ D^\prime=\{d_1^\prime,...,d_n^\prime\}$ \\
         Noisy maximum degree $d_{max}^\prime$\\
      \end{tabular} \\
      
		\ENSURE  %Output 
		Projected adjacent matrix $\hat{A}$

            \FOR{each user $v_i, i\in[1,n]$ in a graph}
            \IF{$d_i > d_{max}^\prime$} 
            \STATE Initialize: $ds=[0]*n, \hat{A_i}=\varnothing$
            \FOR{$j=1$ \textbf{to} $n$}  
            \IF{$A_{ij}==1$}
            \STATE $ds[j] \leftarrow \mathsf{DS}(d_i,d_j^\prime)$
            \ENDIF
            \ENDFOR
            \STATE Sort $ds$ in ascending order
            \STATE $\hat{ds} \leftarrow ds[1:d_{max}^\prime]$
            \FOR{$j=1$ \textbf{to} $n$}       
            \IF{$ds[j]$ in $\hat{ds}$}
            \STATE $\hat{ A_i} \leftarrow \hat{ A_i} \cup \{1\}$
            \ELSE
            \STATE $\hat{ A_i} \leftarrow \hat{ A_i} \cup \{0\}$
            \ENDIF
            \ENDFOR
        \ELSIF{$d_i \leq d_{max}^\prime$}
         \STATE $\hat{A_i} \leftarrow A_i$
         \ENDIF
         \ENDFOR
            \RETURN $\hat{ A}$
	\end{algorithmic} 
\end{algorithm}

\begin{observation} 
%\vspace{-0.5cm}
[Triangle Homogeneity \cite{durak2012degree}]
    \label{lem:degree correlations of a triangle}
    Node degrees of a triangle are quite similar to each other in a graph (i.e., social and interaction graphs).
\end{observation}

Durak \textit{et al.} \cite{durak2012degree} demonstrate this observation through experiments conducted on graphs from various scenarios. 
According to Observation~\ref{lem:degree correlations of a triangle}, a significant number of triangles can be preserved by selecting nodes with a high degree of similarity. 
This is an intuition behind our proposed local graph projection method in triangle counting. 
The degree similarity between two nodes is quantified as outlined in Definition \ref{def:degree similarity}. 
It is important to note that Equation \ref{equation:ds} reflects the relative difference in degrees between two nodes. 
Consequently, a lower value of $DS(d_1,d_2)$ indicates a higher degree of similarity between the nodes.

\begin{definition} 
% \begin{lemma} 
[Degree Similarity]
    \label{def:degree similarity}
    Given two node degrees $d_1$ and $d_2$ in a graph, the degree similarity between them is computed by
    \begin{equation}
    \label{equation:ds}
        DS(d_1,d_2)=\frac{|d_1-d_2|}{d_1}
    \end{equation}
\end{definition}
% \end{lemma}

Algorithm \ref{alg:local projection} shows the details of our local graph projection method.
It takes as input the adjacent matrix of a graph, true degree set $D$, noisy degree set $D^\prime$, and noisy maximum degree $d_{max}^\prime$.
If $d_i > d_{max}^\prime$, each user $v_i$ first initializes an array $ds$ with size $n$ and an empty set $\hat{A_i}$, where $ds$ records degree similarities and $\hat{A_i}$ is the projected adjacent bit vector (line 3).
Then user $v_i$  computes the degree similarities between $d_i$ and all her friends based on Definition \ref{def:degree similarity}, and records the similarities using the array $ds$ (line 5).
Here, the neighboring node degrees are noisy degrees that have been calculated in $\mathsf{Max}$ function (Algorithm \ref{alg:maximum degree}).
Then the array $ds$ is sorted in ascending order (line 6), and then the top $d_{max}^\prime$ elements are sliced and stored into the array $\hat{ds}$ (line 7).
After that, user $v_i$ traverses each neighboring friend $v_k, k\in[n]$ and checks the degree similarity $ds[k]$ between them.
If the element $ds[k]$ is in $\hat{ds}$, the bit `1' will be added into $\hat{A_i}$; 
the bit `0' will be added otherwise (line 9-12).
If $d_i \leq d_{max}^\prime$, the projected adjacent bit vector $\hat{A_i}$ will be set as the original vector $A_i$.
The final answer is the projected adjacent matrix $\hat{A}$.
We denote this algorithm by $\mathsf{Project}$.

\subsection{Additive Secret Sharing-based Triangle Counting}
\label{subsec:local triangle count}
% The intuition of triangle counting is that 
Our triangle counting strategy is based on an intriguing fact that  a triangle exists
if three neighboring edges of a triple exist simultaneously, namely, $a_{ij}\times a_{ik} \times a_{jk}=1$, where $i,j,k \in [1,n]$.
The challenge lies in computing the multiplication of three adjacent bits while preserving the privacy of neighboring information.
We seek help from the multiplication of secret values using additive secret sharing.
However, existing protocols \cite{mohassel2017secureml,riazi2018chameleon} mainly focus on multiplying two secret values, which cannot be directly employed for triangle counting.

\begin{algorithm}[t]
\small
	\caption{$\mathsf{Count}$: ASS-based Triangle Counting}
	\label{alg:local triangle counting} 
	\begin{algorithmic}[1] 
 
	\REQUIRE  %Input 
       \begin{tabular}[t]{l}
         Projected adjacent  matrix $\hat{A}=\{\hat{A_1},...,\hat{A_n}\}$\\
      \end{tabular} \\
      
	\ENSURE  %Output 
	Secret shares of triangle count $\langle T \rangle$

          \STATE Initialize: $\langle T \rangle_1=\langle T \rangle_2=0$
          \FOR{$i=1$ \textbf{to} $n$}
          \FOR{$j=i+1$ \textbf{to} $n$}
          \FOR{$k=j+1$ \textbf{to} $n$}
          \STATE Initialize: \begin{tabular}[t]{l}
             $w=xyz, o=xy,p=xz,q=yz$ \\
            $\langle x \rangle_1,\langle y \rangle_1,\langle z \rangle_1,\langle w \rangle_1,\langle o \rangle_1 ,\langle p \rangle_1 ,\langle q \rangle_1  \rightarrow S_1$\\
            $\langle x \rangle_2,\langle y \rangle_2,\langle z \rangle_2,\langle w \rangle_2,\langle o \rangle_2,\langle p \rangle_2,\langle q \rangle_2 \rightarrow S_2$
            \end{tabular}
        \STATE Server $S_1$:\begin{tabular}[t]{l} 
        $\langle e \rangle_1=\langle a_{ij} \rangle_1- \langle x \rangle_1$ \\
        $\langle f \rangle_1=\langle a_{ik} \rangle_1- \langle y \rangle_1$ \\
        $\langle g \rangle_1=\langle a_{jk} \rangle_1- \langle z \rangle_1$ \\
          \end{tabular}
        \STATE Server $S_2$:\begin{tabular}[t]{l} 
        $\langle e \rangle_2=\langle a_{ij} \rangle_2- \langle x \rangle_2$ \\
        $\langle f \rangle_2=\langle a_{ik} \rangle_2- \langle y \rangle_2$ \\
        $\langle g \rangle_2=\langle a_{jk} \rangle_2- \langle z \rangle_2$ \\
          \end{tabular}
        \STATE Server $S_1$ and $S_2$ communicate and obtain:\\
        \ $e=\langle e \rangle_1+\langle e \rangle_2$, 
        $f=\langle f \rangle_1+\langle f \rangle_2$,
        $g=\langle g \rangle_1+\langle g \rangle_2$
        \STATE Server $S_1$:\begin{tabular}[t]{l}
        $u_1=\langle w \rangle_1+\langle xy \rangle_1 g+\langle xz \rangle_1 f+\langle yz \rangle_1 e$\\
        $+\langle x \rangle_1 fg+\langle y \rangle_1 eg+\langle z \rangle_1 ef$
        \end{tabular}
        \STATE Server $S_2$:\begin{tabular}[t]{l}
        \ $u_2=\langle w \rangle_2+\langle xy \rangle_2 g+\langle xz \rangle_2 f+\langle yz \rangle_2 e$\\
        $+\langle x \rangle_2 fg+\langle y \rangle_2 eg+\langle z \rangle_2ef+efg$
        \end{tabular}
        \STATE Server $S_1$: $\langle T \rangle_1 \leftarrow \langle T \rangle_1 + u_1$
        \STATE Server $S_2$: $\langle T \rangle_2 \leftarrow \langle T \rangle_2 + u_2$
          \ENDFOR
          \ENDFOR
          \ENDFOR
        \STATE $\langle T \rangle \leftarrow \{\langle T \rangle_1, \langle T \rangle_2\}$
        \RETURN $\langle T \rangle$
	\end{algorithmic} 
\end{algorithm}

Motivated by this, we propose a secure multi-party triangle counting protocol that executes the multiplication of three secret values utilizing additive secret sharing.
Given three secret values $a,b,c$, our goal is to compute the multiplication of these three secrets while not leaking anything about secrets, namely, $d=a\times b \times c$.
Like the multiplication of two secrets, two servers precompute Multiplication Groups (MGs) via oblivious transfer \cite{rabin2005exchange,kilian1988founding}.
MGs refer to a set of shared values:$x,y,z,w,o,p,q$, where $w=x \times y \times z,o=x\times y,p=x\times z, q=y\times z$.
Each value is represented as an $l$-bit integer in the ring $\mathbb{Z}_{2^l}$.
In offline phase, server $S_1$ receives $\langle x \rangle_1,\langle y \rangle_1,\langle z \rangle_1, \langle w \rangle_1,\langle o \rangle_1,\langle p \rangle_1,\langle q \rangle_1$, and server $S_2$ receives $\langle x \rangle_2,\langle y \rangle_2,\langle z \rangle_2,\langle w \rangle_2,\langle o \rangle_2,\langle p \rangle_2,\langle q \rangle_2$.
After having shares of MGs, the multiplication is performed as follows:
\begin{enumerate}
    \item Server $S_i$ $(i\in\{1,2\})$ computes $\langle e \rangle_i=\langle a\rangle_i- \langle x \rangle_i,$ \\
        $\langle f \rangle_i=\langle b \rangle_i- \langle y \rangle_i,$ and $\langle g \rangle_i=\langle c \rangle_i- \langle z \rangle_i$ 
    \item Both server $S_1$ and $S_2$ communicate to reconstruct $e,f,$ and $g$.
    \item Server $S_i$ computes its secret share of the multiplication result as: $\langle d \rangle_i=\langle w \rangle_i+\langle xy \rangle_i g+\langle xz \rangle_i f+\langle yz \rangle_i e+\langle x \rangle_i fg+\langle y \rangle_i eg+\langle z \rangle_ief+(i-1)efg$
\end{enumerate}

\begin{theorem}
\label{def:mq correctness}
    Our proposed multiplication of three secret values is correct.
\end{theorem}
\emph{Proof of Theorem \ref{def:mq correctness}}. 
$d=\langle d \rangle_1 + \langle d \rangle_2=\langle w \rangle_1+\langle xy \rangle_1 g+\langle xz \rangle_1 f+\langle yz \rangle_1e+\langle x \rangle_1 fg+\langle y \rangle_1 eg+\langle z \rangle_1ef
+\langle w \rangle_2+\langle xy \rangle_2 g+\langle xz \rangle_2 f+\langle yz \rangle_2 e+\langle x \rangle_2 fg+\langle y \rangle_2 eg+\langle z \rangle_2ef+efg$\\
$=w+xyg+xzf+yze+xfg+yeg+zef+efg$.\\
Then, we put $e=a-x, f=b-y, g=c-z$ into above equation and obtain $d=a\times b \times c$.\qed

Algorithm \ref{alg:local triangle counting} shows how to securely calculate the secret shares of true triangle count based on our proposed multiplication protocol of three secret shares.
It takes as input a projected graph that is represented as an adjacent matrix $\hat{A}=\{\hat{A_1},...,\hat{A_n}\}$.
It traverses all possible triangle triples by computing $u=a_{ij}\times a_{ik} \times a_{jk}$.
If $u=1$, these three nodes and three edges constitute a triangle; otherwise not.
Note that we reduce the repeat computation by only $u=a_{ij}\times a_{ik} \times a_{jk} (i<j<k)$.
Each user secretly shares each bit of its adjacent bit vector to two servers.
Currently, server $S_1$ obtains $\langle a_{ij} \rangle_1$ and $S_2$ owns $\langle a_{ij} \rangle_2$, where $i,j \in[1,n]$.
For each possible triangle, it first initializes the multiplication groups including $x,y,z,w,o,p,q$, where $w=x \times y \times z,o=x\times y,p=x\times z, q=y\times z$.
The multiplication groups can be precomputed in the offline phase and then shared with two servers (line 5).
After that, server $S_1$ computes  $\langle e \rangle_1=\langle a_{ij} \rangle_1- \langle x \rangle_1$, $\langle f \rangle_1=\langle a_{ik} \rangle_1- \langle y \rangle_1$, $\langle g \rangle_1=\langle a_{jk} \rangle_1- \langle z \rangle_1$.
Server $S_2$ computes $\langle e \rangle_2=\langle a_{ij} \rangle_2- \langle x \rangle_2$, $\langle f \rangle_2=\langle a_{ik} \rangle_2- \langle y \rangle_2$, $\langle g \rangle_2=\langle a_{jk} \rangle_2- \langle z \rangle_2$ (line 6-7).
Next, two servers communicate and reconstruct $e,f,$ and $g$.
It is worth noting that secret values are masked with random values $x,y,z$, and thus any server knows nothing about $a_{ij}, a_{ik}, a_{jk}$.
Then, server $S_1$ computes the secret share of the current triangle, namely, $u_1$.
Server $S_2$ computes the secret share of the current triangle, namely, $u_2$ (line 9-10).
$S_1, S_2$ adds $u_1, u_2$ into $\langle T \rangle_1$ and $\langle T \rangle_2$, respectively.
Finally, each server obtains the secret share of the true triangle counting, namely, $\langle T \rangle_1$ and $\langle T \rangle_2$.
The server cannot know any information about true triangle count $T$ from $\langle T \rangle_1$ or $\langle T \rangle_2$.
The final answer of Algorithm \ref{alg:local triangle counting} is the secret shares of true triangles.
We denote this algorithm by $\mathsf{Count}$.

\begin{algorithm}[t]
\small
	\caption{$\mathsf{Perturb}$: Distributed Perturbation} 
	\label{alg:randomization with distributed DP} 
	\begin{algorithmic}[1] 
 
		\REQUIRE  %Input 
       \begin{tabular}[t]{l}
         Secret share of triangle count $\langle T \rangle$,\\
        Noisy maximum degree $d_{max}^\prime$,\\
        Privacy budget $\varepsilon_2$\\
      \end{tabular} \\
      
		\ENSURE  %Output 
		Noisy triangle count $T^\prime$

           \FOR{each user $v_i$, $i\in[1,n]$}
            \STATE $Gam_1=\mathsf{Gamma}(n,\frac{d_{max}^\prime}{\varepsilon_2})$
            \STATE $Gam_2=\mathsf{Gamma}(n,\frac{d_{max}^\prime}{\varepsilon_2})$
            \STATE $\gamma_i=(Gam_1-Gam_2)$
            \STATE Split $\gamma_i$ into two secret shares \\
             $ \gamma_i=\langle \gamma_i \rangle_1+\langle \gamma_i \rangle_2$
            \STATE Send $\langle \gamma_i \rangle_1, \langle \gamma_i \rangle_2$ to two servers $S_1, S_2$\\
             $\langle \gamma_i \rangle_1 \rightarrow S_1$, $\langle \gamma_i \rangle_2 \rightarrow S_2$
           \ENDFOR
            \STATE Server $S_1$: \begin{tabular}[t]{l}
            Aggregate $n$ shared distributed \\
            $\langle \gamma \rangle_1=\sum_{i=1}^n \langle \gamma_i \rangle_1$ 
            \end{tabular}
            \STATE Server $S_2$: \begin{tabular}[t]{l}
            Aggregate $n$ secret shares \\
            $\langle \gamma \rangle_2=\sum_{i=1}^n \langle \gamma_i \rangle_2$ 
            \end{tabular}
            \STATE Server $S_1$: \begin{tabular}[t]{l}
            Compute the secret share of $T^\prime$ \\
            $\langle T^\prime \rangle_1=\langle T \rangle_1+\langle \gamma \rangle_1$
            \end{tabular}
            \STATE Server $S_2$: \begin{tabular}[t]{l}
            Compute the secret share of $T^\prime$ \\
            $\langle T^\prime \rangle_2=\langle T \rangle_2+\langle \gamma \rangle_2$
            \end{tabular}
            \STATE Server $S_1, S_2$: Communicate and compute \\
            $T^\prime=\langle T^\prime \rangle_1+\langle T^\prime \rangle_2$
            \RETURN $T^\prime$
	\end{algorithmic} 
\end{algorithm}

\subsection{Distributed Perturbation}
\label{subsec:randomization via DDP}
Upon computing the true triangle count $T$,  we can add DP noise into the triangle count, in order to guarantee that the final output is differentially private.
The state-of-the-art crypto-assisted differential privacy model \cite{roy2020crypt} adds two instances of Laplace noise to the query result in tabular data analysis.
However, the additional round randomization brings more noise than the CDP model.
We propose a distributed perturbation method by combining the additive secret sharing \cite{shamir1979share,mohassel2017secureml,riazi2018chameleon} and distributed noise generation method \cite{shi2011privacy,acs2011have,goryczka2015comprehensive}.
Each user first generates sufficient but minimal noise and then sends it to two servers. 
Since such a noise is also sensitive,
we employ additive secret sharing to encode the noise and share them with two servers, respectively. 
The server can not obtain any information about each added noise and the final noisy triangle count is protected under $\varepsilon$-Edge Distributed DP.
The Laplace mechanism has been widely used for protecting the triangle counting in CDP or LDP model \cite{ding2021differentially,imola2021locally,imola2022communication}, and a key property of Laplace distribution, namely, infinite divisibility \cite{kotz2001laplace,goryczka2015comprehensive} (as presented in Lemma~\ref{def:divisibility}), allows us to simulate the Laplace noise by summing up $n$ other random variables from independent identically distribution.

\begin{lemma}[Infinite Divisibility \cite{kotz2001laplace}]
	\label{def:divisibility}
	Let $Lap(\lambda)$ denote a random variable that is sampled from a Laplace distribution with \emph{PDF} $f(x,\lambda)=\frac{1}{2\lambda}e^{\frac{|x|}{\lambda}}$. 
    Then the distribution of $Lap(\lambda)$ is infinitely divisible; 
    i.e., $Lap(\lambda)$ can be expressed as the sum of an arbitrary number of independent and identically distributed (i.i.d.) random variables. Specifically,
    % Moreover, 
    for arbitrary integer $n \geq 1$, 
    \begin{equation}
        Lap(\lambda)=\sum_{i=1}^n[Gam_1(n,\lambda)-Gam_2(n,\lambda)],
    \end{equation}
    where $Gam_1(n,\lambda)$ and $Gam_2(n,\lambda)$ are independent Gamma distributed random variables with densities as follows,
    \begin{equation}
        Gamma(x,n,\lambda)=\frac{(1/\lambda)^{1/n}}{\Gamma(1/n)}x^{\frac{1}{n}-1}e^{-\frac{x}{\lambda}},
    \end{equation}
    where $\Gamma$ is the Gamma function, such that $\Gamma(\beta)=\int_{0}^{\infty}x^{\beta-1}e^{-x}dx$.
\end{lemma}

Algorithm \ref{alg:randomization with distributed DP} contains the details of perturbing the triangle count using distributed noise.
It takes as input the encoded triangle count $\langle T \rangle=\{\langle T \rangle_1, \langle T \rangle_2\}$, noisy maximum degree $d_{max}^\prime$, and privacy budget $\varepsilon_2$.
According to Lemma \ref{def:divisibility}, one $Lap(.)$ random variable can be obtained by summing up $2n$ random variables.
Each user $v_i$ samples two random variables $Gam_1$ and $Gam_2$ from $\mathsf{Gamma}$ distribution, and then obtains a partial noise by computing $\gamma_i=(Gam_1-Gam_2)$ (line 2-4).
To avoid the size of noise, each user encodes $\gamma_i$ using additive secret sharing, i.e., $ \gamma_i=\langle \gamma_i \rangle_1+\langle \gamma_i \rangle_2$,  and then sends the shares to two servers (line 5-6).
Server $S_1, S_2$ collects all secret shares of the partial noise and aggregates them as: $\langle \gamma \rangle_1=\sum_{i=1}^n \langle \gamma_i \rangle_1$ and $\langle \gamma \rangle_2=\sum_{i=1}^n \langle \gamma_i \rangle_2$, respectively (line 7-8).
Each server adds the secret share of noise into the share of triangle count, namely, $\langle T^\prime \rangle_1=\langle T \rangle_1+\langle \gamma \rangle_1$ and $\langle T^\prime \rangle_2=\langle T \rangle_2+\langle \gamma \rangle_2$ (line 9-10).
After communicating with each other, server $S_1, S_2$ can obtain the final noisy triangle counting result $T^\prime$.
This algorithm is denoted by $\mathsf{Perturb}$.

\section{Theoretical Analysis}
\label{sec:theoretical analysis}
\subsection{Security and Privacy Analysis}

\subsubsection{Security Analysis}
We first analyze the security of our proposed Algorithm \ref{alg:local triangle counting} and Algorithm \ref{alg:randomization with distributed DP}.
Following the simulation-based paradigm \cite{lindell2017simulate}, we prove the security guarantee by giving a simulator, ensuring that simulator's view and each server's real view are computationally indistinguishable.

\begin{definition}
\label{def:instinguishability}
    Let $\Pi$ denote the protocol in the semi-honest and non-colluding scenario.
    Let $\mathsf{view}_{S_i}^{\Pi}$ be the view of the server $S_i$.
    Let $\mathsf{Sim}_{S_i}$ be the view of a simulator.
    The execution of $\Pi$ is secure if $\mathsf{view}_{S_i}^{\Pi}$ and $\mathsf{Sim}_{S_i}$ are computationally indistinguishable, namely, $\mathsf{view}_{S_i}^{\Pi} \approx \mathsf{Sim}_{S_i}$.
\end{definition}

\begin{theorem}
	\label{theorem:count_security count}
 Given the security of additive secret sharing, our Algorithm \ref{alg:local triangle counting} and Algorithm \ref{alg:randomization with distributed DP} are secure according to Definition \ref{def:instinguishability}.
\end{theorem}

For multiplication in Algorithm \ref{alg:local triangle counting} and addition in Algorithm~\ref{alg:randomization with distributed DP} via additive secret sharing, the random split of secret values guarantees that both the simulator view and the real view are identical.
Each server obtains no information about user's sensitive information, where the information leakage from aggregation result is bounded and qualified via formal DP (refer to \emph{Privacy Analysis}).

\subsubsection{Privacy Analysis}
Then, we present privacy guarantee of Algorithm \ref{alg:maximum degree} and Algorithm \ref{alg:overall protocol}.

\begin{theorem}
	\label{theorem:privacy_max}
	Algorithm \ref{alg:maximum degree} satisfies $\varepsilon_1$-Edge LDP.
\end{theorem}

\emph{Proof of Theorem \ref{theorem:privacy_max}}. 
Let $A_i$ and $A_i^\prime$ be two neighboring adjacent lists that differ in one edge, and $d_i$ and $d_i^\prime$ be their node degrees respectively.
Clearly, $|d_i - d_i^\prime| = 1$.
Let the noise $x$ and $x^\prime$ are two random values drawn from $Lap(\frac{1}{\varepsilon_1})$, the probability of outputting the same noisy degree $d^\prime$ can be bounded by:
{\small
\begin{align*}
     & \frac{Pr[d^\prime=d_i+x]}{Pr[d^\prime=d_i^\prime+x^\prime]}=\frac{Pr[x=d^\prime-d_i]}{Pr[x^\prime=d^\prime-d_i^\prime]} \\
     = & \frac{e^{-\varepsilon_1.|d^\prime-d_i|}}{e^{-\varepsilon_1.|d^\prime-d_i^\prime|}} 
     = e^{\varepsilon_1.(|d^\prime-d_i^\prime|-|d^\prime-d_i|)} 
     \leq  e^{\varepsilon_1|d_i-d_i^\prime|}=
     e^{\varepsilon_1},
     %e^{\varepsilon_1}
\end{align*}
}which proves that $d_i$ satisfies 
$\varepsilon_1$-Edge LDP.
% $\varepsilon_1$-Relationship DP.
Then, according to the post-processing property \cite{dwork2014algorithmic}, Algorithm \ref{alg:maximum degree} satisfies 
$\varepsilon_1$-Edge LDP.\qed
% $\varepsilon_1$-Relationship DP.
% Based on Proposition \ref{proposition:edgeldp_relationship}, we can further prove that Algorithm \ref{alg:maximum degree} satisfies 
% $\varepsilon_1$-Relationship DP.\qed
% $\frac{1}{2}\varepsilon_1$-Edge LDP.

\begin{theorem}
\label{theorem:cargo_ddp}
    Algorithm \ref{alg:overall protocol} satisfies 
    %$\varepsilon_2$-
    ($\varepsilon_1 + \varepsilon_2$)-Edge DDP (Definition \ref{def:Edge DDP}).
\end{theorem}

\begin{table*}[t]
\small
	\caption{Summary of Theoretical Results.}
	\centering
	\label{tab:theoretical comparison}
	\setlength{\tabcolsep}{2mm}{
		\begin{tabular}{|c|c|c|c|}
			\hline
&$\mathsf{CentralLap_\triangle}$  & CARGO& $\mathsf{Local2Rounds_\triangle}$  \\\hline
Server & Trusted & Untrusted & Untrusted \\\hline
Privacy & $\varepsilon$-Edge CDP & ($\varepsilon_1+\varepsilon_2$)-Edge DDP & $\varepsilon$-Edge LDP\\\hline
% Privacy & $\varepsilon$-Edge CDP & $\varepsilon_2$-Edge DDP & $\varepsilon$-Edge LDP\\\hline
Utility & $O(\frac{d_{max}^2}{\varepsilon^2})$ & $O(\frac{d_{max}^{\prime2}}{\varepsilon_2^2})$ & $O(\frac{e^\varepsilon}{(e^\varepsilon-1)^2}(d_{max}^3n+\frac{e^\varepsilon}{\varepsilon^2}d_{max}^2n))$ 
\\\hline       
Time Complexity& $O(1)$ & $O(n^3)$ & $O(n^2+nd_{max}^2)$\\\hline
	\end{tabular}}
\end{table*}

\emph{Proof of Theorem \ref{theorem:cargo_ddp}}. 
Algorithm \ref{alg:overall protocol} uses two privacy budgets: $\varepsilon_1$ in $\mathsf{Max}$ and $\varepsilon_2$ in $\mathsf{Perturb}$. 
By Theorem~\ref{theorem:privacy_max}, $\mathsf{Max}$ provides $\varepsilon_1$-Edge LDP. 
Below, we analyze the privacy of $\mathsf{Perturb}$. 

Let $G$ and $G^\prime$ be two neighboring graphs that differ in one edge, and $T(G)$ and $T(G^\prime)$ are their corresponding triangle counts respectively.
The sensitivity of triangle counting is denoted by $|T(G)-T(G^\prime)|=\triangle$.
Let $\langle T \rangle_1$ and $\langle T \rangle_2$ be two secret shares for true triangle count $T$.
Then, we have
\begin{center}
    $T(G)=\langle T \rangle_1+\langle T \rangle_2$, $T(G^\prime)=\langle T^\prime \rangle_1+\langle T^\prime \rangle_2$
\end{center}
Let $r=\{r_1,...,r_n\}$ and  $r^\prime=\{r_1^\prime,...,r_n^\prime\}$ be two sets of distributed noise from 
$Gam_1(n,\frac{\triangle}{\varepsilon_2})-Gam_2(n,\frac{\triangle}{\varepsilon_2})$. 
Let $x$ and $x^\prime$ be two random variables sampled from $Lap(\frac{\triangle}{\varepsilon_2})$ distribution.
According to the infinite divisibility of $Laplace$ distribution (Lemma \ref{def:divisibility}),  $x$ and $x^\prime$ can be denoted by:
\begin{center}
\small
    $x=r_1+...+r_n$,
    $x^\prime=r_1^\prime+...+r_n^\prime$
\end{center}
The probability of outputting the same noisy triangle count $\widetilde{T}$  can be bounded by:
{\small
\begin{align*}
    &\frac{Pr[\widetilde{T}=\langle T \rangle_1+\langle T \rangle_2+(r_1+...+r_n)]}{Pr[\widetilde{T}=\langle T^\prime \rangle_1+\langle T^\prime \rangle_2+(r_1^\prime+...+r_n^\prime)]} \\
    =&\frac{Pr[\widetilde{T}=T+(r_1+...+r_n)]}{Pr[\widetilde{T}=T^\prime+(r_1^\prime+...+r_n^\prime)]}  \\
    =&\frac{Pr[\widetilde{T}=T(G)+x]}{Pr[\widetilde{T}=T(G^\prime)+x^\prime]}=\frac{Pr[x=\widetilde{T}-T(G)]}{Pr[x^\prime=\widetilde{T}-T(G^\prime)]}\\
    =& \frac{e^{\frac{-\varepsilon_2.|\widetilde{T}-T(G)|}{\triangle}}}{e^{\frac{-\varepsilon_2.|\widetilde{T}-T(G^\prime)|}{\triangle}}}
    = e^\frac{\varepsilon_2.(|\widetilde{T}-T(G^\prime)|-|\widetilde{T}-T(G)|)}{\triangle} \\
    \leq & e^\frac{\varepsilon_2|T(G)-T(G^\prime)|}{\triangle}=e^{\varepsilon_2}
\end{align*}
}Thus, $\mathsf{Perturb}$ provides $\varepsilon_2$-Edge DDP. 
As described in Section~\ref{subsec:DP_Graphs}, both $\varepsilon$-Edge LDP and $\varepsilon$-Edge DDP protect one edge with privacy budget $\varepsilon$. 
Thus, the entire process of Algorithm \ref{alg:overall protocol} provides ($\varepsilon_1 + \varepsilon_2$)-Edge DDP.
\qed

\subsection{Utility Analysis}
The utility error of CARGO is mainly from the projection loss in Algorithm \ref{alg:local projection} and perturbation error in Algorithm \ref{alg:randomization with distributed DP}.

\begin{theorem}[Projection Loss]
Let $T(G)$ and $\hat{T}(G,d_{max}^\prime)$ be the triangle count before and after projection respectively.
The projection parameter is set as the noisy maximum degree $d_{max}^\prime$.
Then, for any $d_{max}^\prime\geq 0$ and $G$,
    \begin{center}
        $\mathbb{E}[l_2^2(T(G), \hat{T}(G,d_{max}^\prime))]=(T(G)-\hat{T}(G,d_{max}^\prime))^2$
    \end{center}
\end{theorem}

\begin{theorem}[Perturbation Error]
\label{theorem:perturbation error}
    \begin{flushleft}
    Let $T^\prime(G,\varepsilon_2,d_{max}^\prime)$ and  
    \end{flushleft}
    \noindent$\hat{T}(G,d_{max}^\prime)$ be the triangle count before and after perturbation, respectively.
    Then, for any privacy budget $\varepsilon_2\geq 0$, noisy maximum degree $d_{max}^\prime\geq 0$, and graph $G$,
    \begin{center}
        $\mathbb{E}[l_2^2(T^\prime(G,\varepsilon_2,d_{max}^\prime),\hat{T}(G,d_{max}^\prime))]=O(\frac{d_{max}^{\prime2}}{\varepsilon_2^2})$
    \end{center}
\end{theorem}

\emph{Proof of Theorem \ref{theorem:perturbation error}}.
According to the well-known bias-variance decomposition \cite{ murphy2012machine}, the expected $l_2$ loss consists of the bias and variance, which can be written as follows:
\begin{align*}
    & \mathbb{E}[l_2^2(T^\prime(G,\varepsilon_2,d_{max}^\prime),\hat{T}(G,d_{max}^\prime))] \\
    = & (\mathbb{E}[T^\prime(G,\varepsilon_2,d_{max}^\prime)]-\hat{T}(G,d_{max}^\prime))^2+\mathbb{V}[T^\prime(G,\varepsilon_2,d_{max}^\prime)]
\end{align*}
Since the mean of Laplacian noise $Lap(\frac{\triangle}{\varepsilon_2})$ is 0, the estimation $T^\prime(G,\varepsilon_2,d_{max}^\prime)$ is unbiased.
Then, the expected $l_2$ loss will be equal to the variance, which can be formalized as:
\begin{align*}
    &\mathbb{E}[l_2^2(T^\prime(G,\varepsilon_2,d_{max}^\prime),\hat{T}(G,d_{max}^\prime))] 
    = \mathbb{V}[T^\prime(G,\varepsilon_2,d_{max}^\prime)] \\
    = &\mathbb{V}[ Lap(\frac{d_{max}^\prime}{\varepsilon_2})]
    =  O(\frac{d_{max}^{\prime2}}{\varepsilon_2^2})  \tag*{\qed}
\end{align*}

In fact, after empirical analysis, we find that $d_{max}^\prime \approx d_{max}$, as presented in Table \ref{tab:d_max_prime}. 
Specifically, we find that $l_2^2(d_{max}^\prime,d_{max})< 0.009 d_{max}$, 
where $d_{max}$ is the true maximum degree.
% And most 
In addition, $d_{max}^\prime \geq d_{max}$ holds in most cases, which means that there is no deletion of edges during a projection and $\mathbb{E}[l_2^2(T(G), \hat{T}(G,d_{max}^\prime))]=0$.
For a convenient comparison, we omit the projection loss in theoretical analysis. 
Thus, our CARGO attains the expected $l_2$ loss of $O(\frac{d_{max}^{\prime2}}{\varepsilon_2^2})$, where $\varepsilon_2=0.9\varepsilon$ in our setting.

\emph{Discussion}.
It is worth noting that the private maximum degree is the upper bound of local sensitivity, and there are some pros and cons.
On the one hand, it provides a privacy guarantee for the worst case at the expense of utility.
For instance, if we only consider bipartite graphs, the result of triangle counting is always 0.
The local sensitivity is 0 while the maximum degree can be very large, which leads to much error.
Recently, some works \cite{dong2022nearly,dong2021residual} utilize the smooth sensitivity (SS) and residual sensitivity (RS) for common graph queries, which can achieve constant noise.
Furthermore, we compare the value of $d_{max}^\prime$ with values of SS and RS in Table 1 in \cite{dong2022nearly}, as presented in Table III.
We find that $d_{max}^\prime$ can be larger than SS and RS in some graphs, for example, CondMat and HepTh, which means that our approach can add much more noise than SS and RS.
On the other hand, our $d_{max}^\prime$ uses the Laplace distribution and the expected $l_2$ loss (= variance) is finite.
However, considering that RS can be regarded as an instantiation of SS \cite{dong2022nearly}, both SS and RS draw noise from a Cauchy distribution, which has an infinite variance.

\subsection{Time Complexity}
Let $n$ and $d_{max}$ denote the number of users and the maximum degree, respectively.
In CARGO, the time complexity of $\mathsf{Max}$ in Algorithm \ref{alg:maximum degree} is $O(n)$.
Then, the time complexity of $\mathsf{project}$ in Algorithm \ref{alg:local projection} is $O(nd_{max})$ since one user needs to compute degree similarities (Lemma \ref{def:degree similarity}) between her and her all neighboring users.
Next, the time complexity of $\mathsf{Count}$ in Algorithm \ref{alg:local triangle counting} is $O(n^3)$.
This is because CARGO needs to traverse all triples to justify if three edges of a triple exist simultaneously.
Finally, the time complexity of $\mathsf{Perturb}$ in Algorithm \ref{alg:randomization with distributed DP} is $O(n)$, and the time complexity of secret sharing and aggregation is also $O(n)$.
Thus, the time complexity of our CARGO is $O(n^3)$.

% \vspace{0.5em}
\noindent\textbf{Summary of Theoretical Results}.
Table \ref{tab:theoretical comparison} summarizes the theoretical results compared with the state-of-the-art methods, namely, $\mathsf{CentralLap_\triangle}$ \cite{imola2021locally} and $\mathsf{Local2Rounds_\triangle}$ \cite{imola2021locally}.
The results of $\mathsf{CentralLap_\triangle}$ and $\mathsf{Local2Rounds_\triangle}$ have been analyzed (refer to Table 2 in \cite{imola2021locally}).
On the one hand, CARGO can achieve a high-utility triangle count comparable to $\mathsf{CentralLap_\triangle}$, but it does not necessarily have a trusted server.
On the other hand, CARGO outperforms much better than $\mathsf{Local2Rounds_\triangle}$ in terms of utility.

\begin{table}[t]
\label{tab:compare sensitivity}
\small
	\caption{Comparison between SS, RS, and $d_{max}^\prime$.}
	\centering
	\setlength{\tabcolsep}{1.5mm}{
		\begin{tabular}{cccccc}
			\hline
            Graph & CondMat &AstroPh&HepPh&HepTh&GrQc\\\hline
		$d_{max}^\prime$ & 560 & 1,008 & 984 & 130 & 162 \\
		SS  & 489&	1,050&	1,350& 102& 183 \\
		RS & 493&	1,054&	1,354&	205& 222\\
   \hline
	\end{tabular}}
 \begin{tablenotes}  
        \footnotesize              
        \item[1] Privacy budget $\varepsilon=1$ 
      \end{tablenotes}          
\end{table}

\section{Experimental Evaluation}
\label{sec:experiment}
In this section, we conduct experiments to answer the following questions:
\begin{itemize}
  \item \textbf{Q1}: What is the utility-privacy trade-off of our CARGO compared with the state-of-the-art CDP-based and LDP-based protocols?
  \item \textbf{Q2}: How does our similarity-based projection method ($\mathsf{Project}$ algorithm) outperform the existing projection method?
  \item \textbf{Q3}: How much running time does our CARGO take compared with competitors?
\end{itemize}

\subsection{Experimental Setting}

\begin{figure*}[t]
	\centering
	\subfigure[Facebook]{
		\begin{minipage}[t]{0.238\linewidth}
			\centering
			\includegraphics[width=\linewidth]{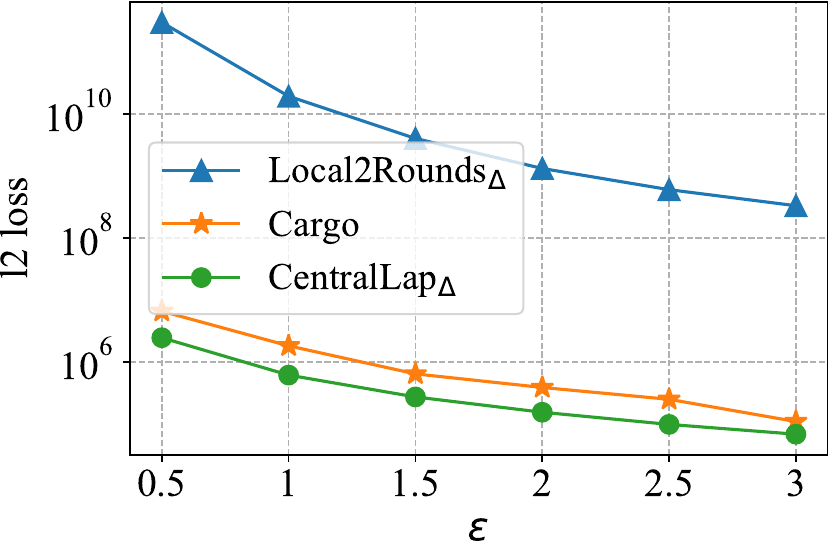}
            \label{fig:l2loss_facebook}
		\end{minipage}
	}%
	\subfigure[Wiki]{
		\begin{minipage}[t]{0.238\linewidth}
			\centering
			\includegraphics[width=\linewidth]{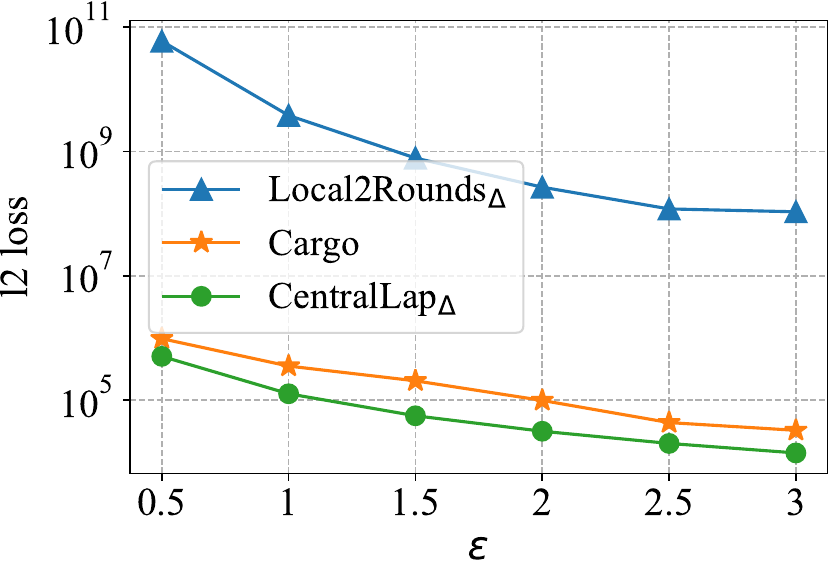}
		\end{minipage}%
	}
 	\subfigure[HepPh]{
		\begin{minipage}[t]{0.238\linewidth}
			\centering
			\includegraphics[width=\linewidth]{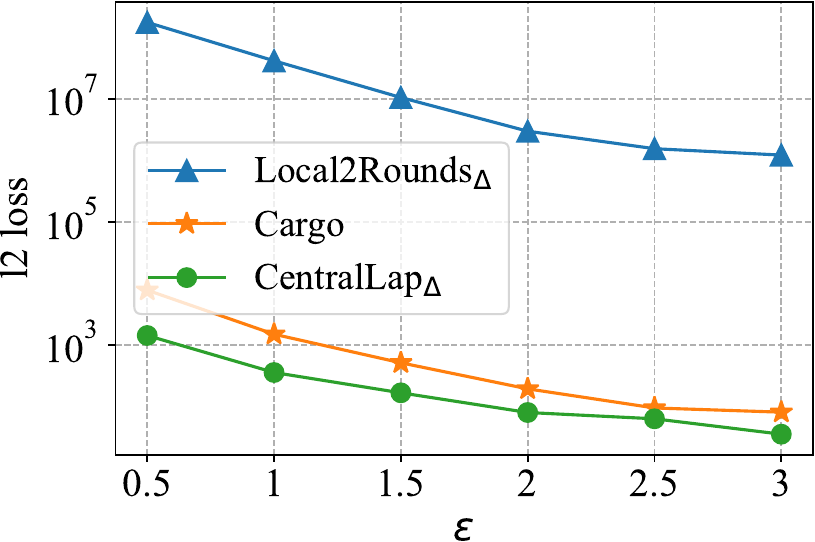}
            \label{fig:l2loss_hepph}
		\end{minipage}%
	}
	\subfigure[Enron]{
		\begin{minipage}[t]{0.238\linewidth}
			\centering
			\includegraphics[width=\linewidth]{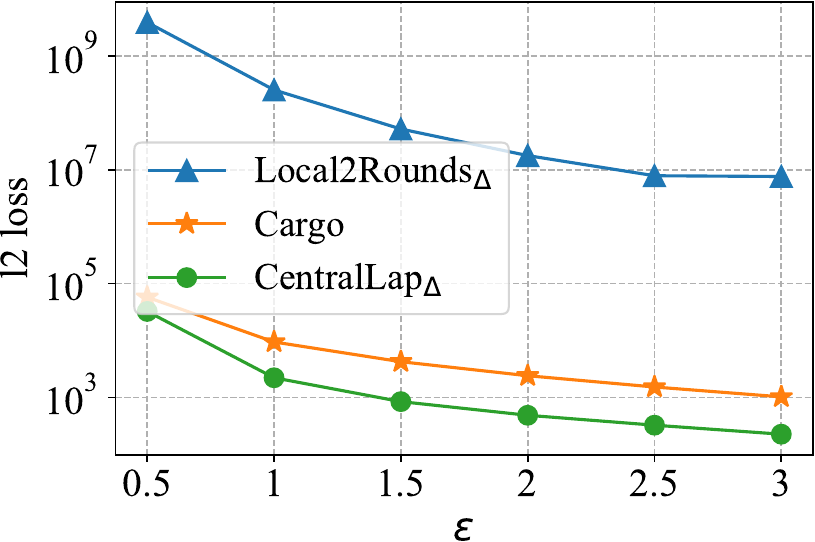}
		\end{minipage}%
	}
\centering
   \vspace{-0.5cm}
\caption{The $l_2$ loss of triangle counting with $\varepsilon$ varying from 0.5 to 3.}
\label{fig:l2loss}
   \vspace{-0.2cm}
\end{figure*}

\begin{figure*}[t]
	\centering
	\subfigure[Facebook]{
		\begin{minipage}[t]{0.238\linewidth}
			\centering
			\includegraphics[width=\linewidth]{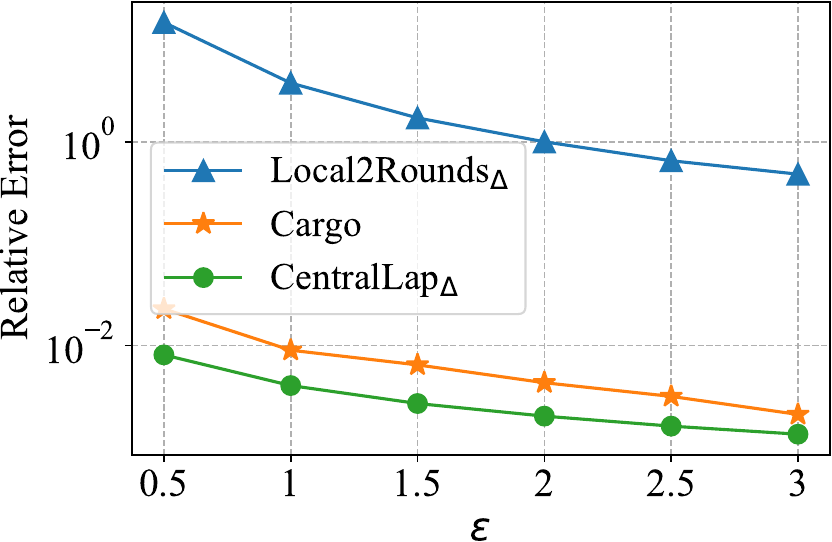}
            \label{fig:re_facebook}
		\end{minipage}
	}%
	\subfigure[Wiki]{
		\begin{minipage}[t]{0.238\linewidth}
			\centering
			\includegraphics[width=\linewidth]{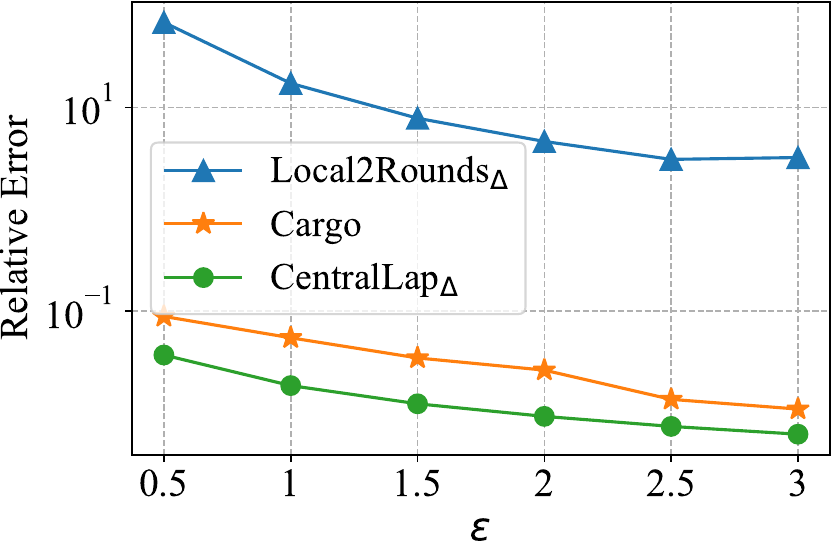}
		\end{minipage}%
	}
 	\subfigure[HepPh]{
		\begin{minipage}[t]{0.238\linewidth}
			\centering
			\includegraphics[width=\linewidth]{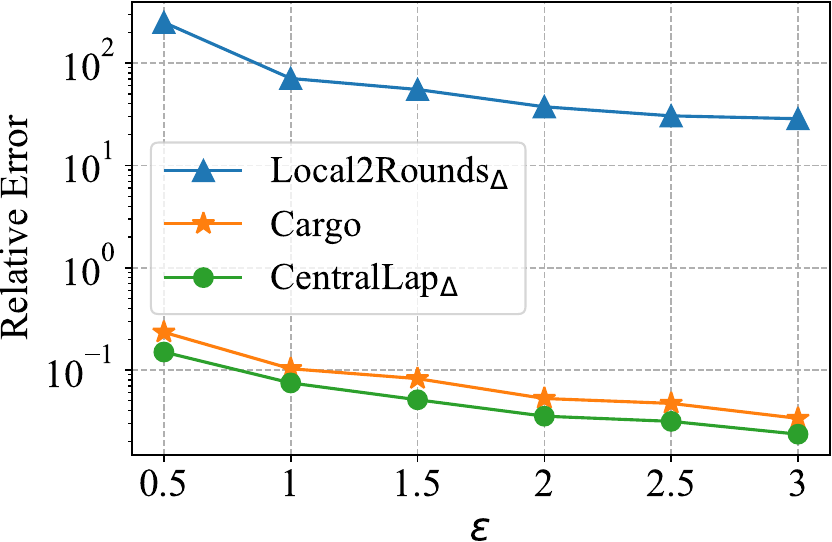}
            \label{fig:re_hepph}
		\end{minipage}%
	}
	\subfigure[Enron]{
		\begin{minipage}[t]{0.238\linewidth}
			\centering
			\includegraphics[width=\linewidth]{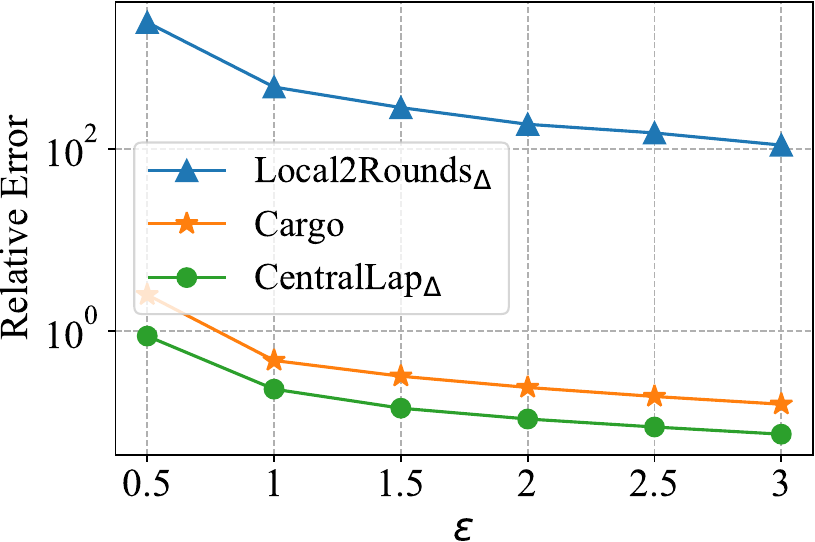}
		\end{minipage}%
	}
    \vspace{-0.5cm}
\centering
\caption{The relative error of triangle counting with $\varepsilon$ varying from 0.5 to 3.}
\label{fig:re}
   \vspace{-0.2cm}
\end{figure*}

\begin{figure*}[t]
 \begin{center}
     \begin{minipage}[t]{0.49\linewidth}
   \centering
\subfigure[Facebook]{
		\begin{minipage}[t]{0.49\linewidth}
			\centering
			\includegraphics[width=\linewidth]{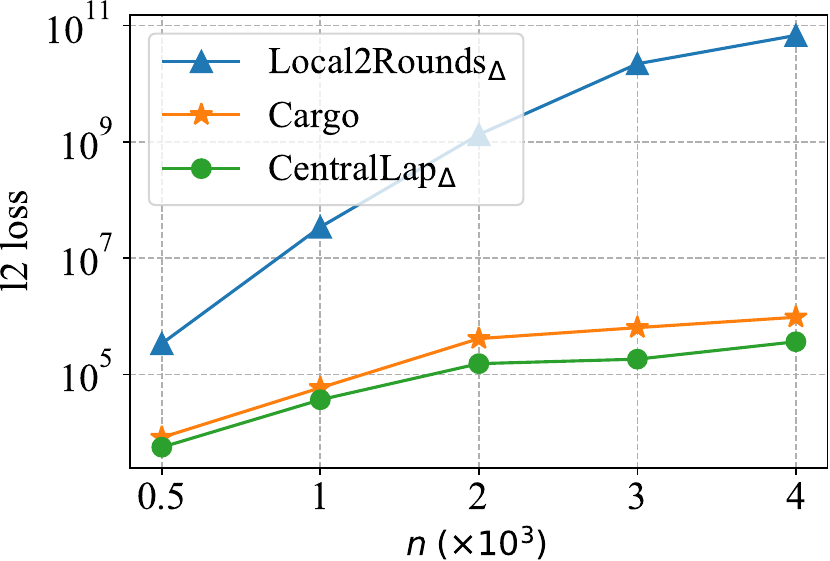}
		\end{minipage}
  \label{fig:l2loss_facebook_n}
	}%
\subfigure[Wiki]{
		\begin{minipage}[t]{0.49\linewidth}
			\centering
			\includegraphics[width=\linewidth]{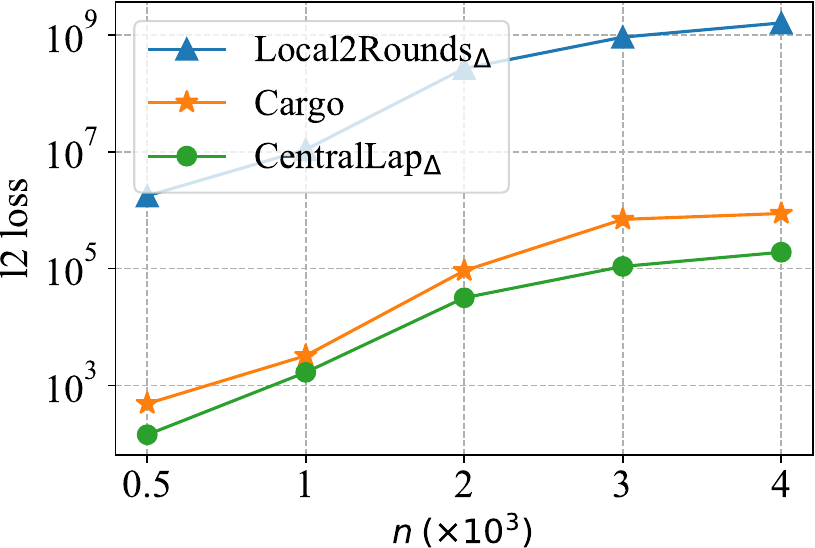}
		\end{minipage}%
	}
    % \vspace{-0.4cm}
       \caption{The $l_2$ loss of triangle counting with different $n$.}
       \label{fig:l2loss_n}
\end{minipage}
 \begin{minipage}[t]{0.49\linewidth}
   \centering
\subfigure[Facebook]{
		\begin{minipage}[t]{0.49\linewidth}
			\centering
			\includegraphics[width=\linewidth]{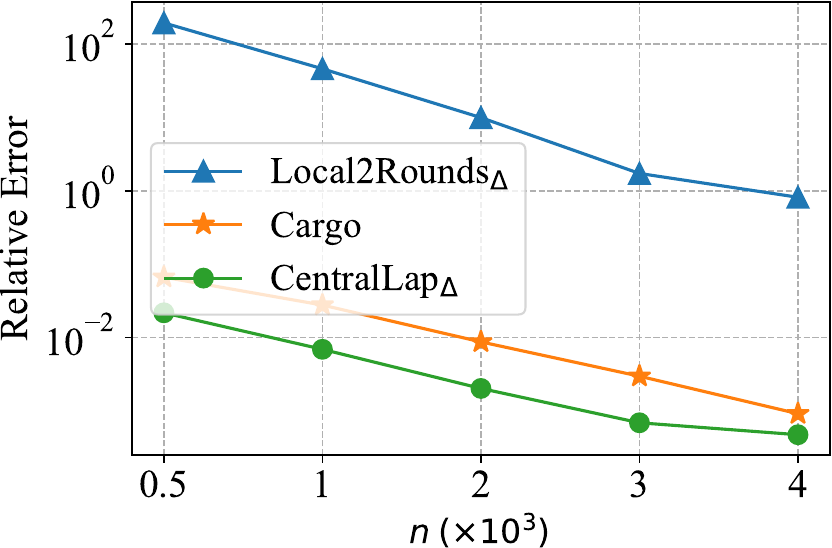}
		\end{minipage}
	}%
\subfigure[Wiki]{
		\begin{minipage}[t]{0.49\linewidth}
			\centering
			\includegraphics[width=\linewidth]{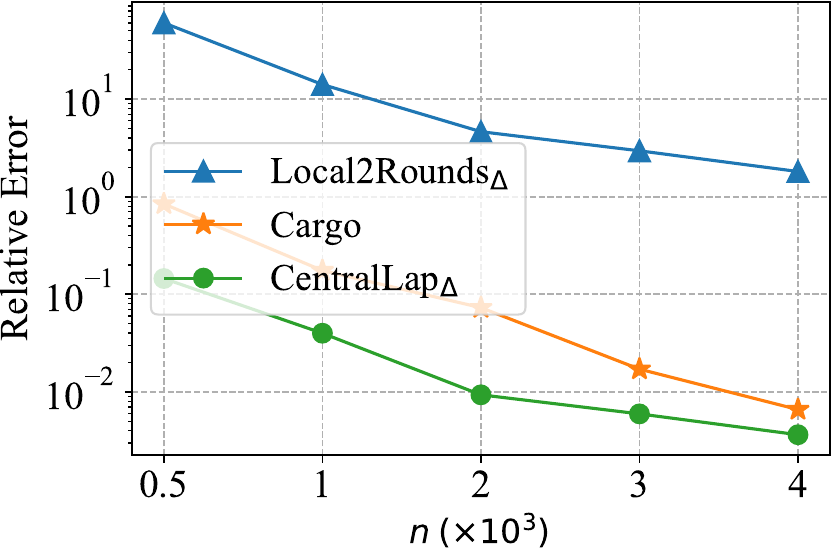}
		\end{minipage}%
  \label{fig:re_wiki_n}
	}
     % \vspace{-0.4cm}
       \caption{The relative error of triangle counting with different $n$.}
       \label{fig:re_n}
\end{minipage}
   \end{center}
      \vspace{-0.5cm}
\end{figure*}

We use four real-world graph datasets from SNAP \cite{snapnets}, and all graphs are preprocessed into undirected and symmetric graphs.
These graphs are from different domains and have different scales.
Table \ref{tab:datasets} presents more details about each graph $G$, including the number of nodes $|V|$, the number of edges $|E|$, the true maximum degree $d_{max}$, and domains that graphs belong to.
For each algorithm, we evaluate the $l_2$ loss and relative error while varying the number of users $n$ and privacy budget $\varepsilon$. 
The default values of $n$ and $\varepsilon$ are $2\times 10^3$ and $2$, respectively. 
Usually, triangle counting needs more privacy budget than the other information (i.e., $d_{max}$) \cite{imola2021locally,sun2019analyzing}.
Thus, we set $\varepsilon_1=0.1\varepsilon$ for publishing noisy maximum degree $d_{max}^\prime$ and $\varepsilon_2=0.9\varepsilon$ for perturbing triangles.

\textbf{Competitors.}
To justify the utility-privacy tradeoff of our CARGO system, we compare it with the state-of-the-art CDP and LDP methods:
(1) $\mathsf{CentralLap_\triangle}$\cite{imola2021locally}, an algorithm for triangle counting using $\mathsf{Laplace}$ mechanism in CDP.
(2) $\mathsf{Local2Rounds_\triangle}$\cite{imola2021locally}, a two-round algorithm for counting triangles in LDP.
To verify the performance of our local projection algorithm (i.e., $\mathsf{Project}$ in Section \ref{subsec:projection}), we compare it with the existing projection method in LDP (i.e., $\mathsf{GraphProjection}$ in \cite{imola2021locally}).

\subsection{Experimental Results}

\textbf{Utility-privacy trade-off}.
We evaluate \textbf{Q1} by comparing the accuracy of our CARGO with that of the aforementioned state-of-the-art LDP and CDP methods when the privacy budget $\varepsilon$ varies from 0.5 to 3.
We also evaluate the accuracy of our $\mathsf{Max}$ algorithm in Section \ref{subsec:projection} that privately estimates the noisy maximum degree $d_{max}^\prime$.
As shown in Table \ref{tab:d_max_prime}, $d_{max}^\prime$ approaches the true maximum degree $d_{max}$ in Table \ref{tab:datasets}.
For each graph, the average of relative error between $d_{max}$ and $d_{max}^\prime$ is less than 1\%.
Thus, we allow an additional round to estimate $d_{max}^\prime$ in our experiments.

Fig. \ref{fig:l2loss} and Fig. \ref{fig:re} present that CARGO significantly outperforms $\mathsf{Local2Rounds_\triangle}$ in all cases.
For example, Fig. \ref{fig:l2loss_facebook} shows that for CARGO, $\varepsilon=3$ results in $l_2$ loss of 1.09$\times 10^5$ as compared to an error of 3.33$\times 10^{8}$ achieved by $\mathsf{Local2Rounds_\triangle}$.
Fig. \ref{fig:l2loss_hepph} illustrates that for $\varepsilon=3$, CARGO owns a $l_2$ loss of 82 while $\mathsf{Local2Rounds_\triangle}$ has an error of 1.22$\times10^{6}$.
Similarly, in Fig. \ref{fig:re_facebook}, CARGO gives a relative error of only 2.11$\times 10^{-3}$ when $\varepsilon=3$. 
In contrast, $\mathsf{Local2Rounds_\triangle}$ has a relative error of 0.48.
Fig. \ref{fig:re_hepph} shows that $\mathsf{Local2Rounds_\triangle}$ has a relative error of 28.6 while CARGO only has an error of 3.37$\times10^{-2}$ for $\varepsilon=3$.
Thus, CARGO improves the $\mathsf{Local2Rounds_\triangle}$ significantly.

Another observation is that the error of CARGO is around $1\times\sim2\times$ larger than that of  $\mathsf{CentralLap_\triangle}$ for most of cases.
With the increase of $\varepsilon$, the error gap between CARGO and $\mathsf{CentralLap_\triangle}$ roughly decreases.
For example, in Fig. \ref{fig:re_facebook}, CARGO has a relative error of 2.29$\times10^{-2}$ as $\mathsf{CentralLap_\triangle}$ has an error of 8.10$\times10^{-3}$ when $\varepsilon=0.5$.
Similarly, when $\varepsilon=3$, Fig. \ref{fig:re_facebook} also shows that the relative error of CARGO is 2.11$\times10^{-3}$ when $\mathsf{CentralLap_\triangle}$ owns a relative error of 1.35$\times10^{-3}$.
This is intuitive because an additional round for privately estimating $d_{max}$ ($\mathsf{Max}$ algorithm) leads to a little error (as shown in Table \ref{tab:d_max_prime}), influencing the overall utility.
On the other hand, increasing $\varepsilon$ improves the accuracy of $\mathsf{Max}$.
Therefore, the error of CARGO is not significantly larger than that of $\mathsf{CentralLap_\triangle}$.

Fig. \ref{fig:l2loss_n} and  Fig. \ref{fig:re_n} show that CARGO outperforms $\mathsf{Local2Rounds_\triangle}$ and achieves the high accuracy comparable to $\mathsf{CentralLap_\triangle}$ while varying $n$ ($\varepsilon=2$).
Here, we just present the results of two graphs due to the limited space.
To be specific, in Fig. \ref{fig:l2loss_facebook_n}, the $l_2$ loss of CARGO is 9.68$\times 10^5$ when $n=4000$. 
In contrast, $\mathsf{Local2Rounds_\triangle}$ owns an $l_2$ loss of 6.8 $\times 10^{10}$.
Furthermore, in Fig. \ref{fig:re_wiki_n}, CARGO has a relative error of 6.59$\times10^{-3}$ when $\mathsf{CentralLap_\triangle}$ has an error of 3.64$\times10^{-3}$ for $n=4000$.
Therefore, CARGO significantly outperforms $\mathsf{Local2Rounds_\triangle}$ and performs similarly to $\mathsf{CentralLap_\triangle}$.

\textbf{Local graph projection}.
Next, we evaluate \textbf{Q2} by performing a comparative analysis between our local projection method ($\mathsf{Project}$) and the existing projection method in local settings ($\mathsf{GraphProjection}$).
We set projection parameter $\theta$ from 10 to 1000, and compute the projection loss by comparing triangle counts before and after projection.
Fig. \ref{fig:proloss_l2loss} and Fig. \ref{fig:proloss_re} show that $\mathsf{Project}$ owns better utility than the baseline $\mathsf{GraphProjection}$ in all cases.
On the other hand, when $\theta$ increases, the projection loss for both of them decreases and the improvement of $\mathsf{Project}$ becomes more significant.
For example, in Fig. \ref{fig:proloss_l2loss_facebook}, the $l_2$ loss of $\mathsf{GraphProjection}$ is 1.01$\times$ larger than that of $\mathsf{Project}$ when $\theta=10$;
In contrast, for $\theta=1000$, the $l_2$ loss of $\mathsf{Project}$ is at least 8$\times$ less than that of $\mathsf{GraphProjection}$.
This is roughly consistent with our theoretical analysis in Section \ref{subsec:projection}.
Namely, randomly deleting edges in $\mathsf{GraphProjection}$ is likely to remove some key edges that involve in many triangles, losing many triangles in a graph.

\begin{table}[t]
% \small
	\caption{details of graph datasets.}
	\centering
	\setlength{\tabcolsep}{1mm}{
		\begin{tabular}{lrrrc}
			\hline
			Graph & $|V|$  & $ |E|$  &$d_{max}$ & Domain\\\hline
			  Facebook & 4,039 & 88,234  & 1,045 & social network \\
			Wiki  & 7,115 & 103,689  & 1,167 & vote network\\
               HepPh & 12,008 & 118,521 & 982 & citation network\\
			Enron & 36,692 & 183,831 & 2,766 & communication network\\
			\hline
	\end{tabular}}
    %\vspace{-0.5cm}
    	\label{tab:datasets}
\end{table}

\begin{table}[t]
\small
	\caption{Noisy maximum degrees under various $\varepsilon$.}
	\centering
	\label{tab:d_max_prime}
	\setlength{\tabcolsep}{2mm}{
		\begin{tabular}{lrrrrrr}
			\hline
	\diagbox[width=5em]{Graph}{\boldsymbol{$\varepsilon$}} & 0.5  & 1 & 1.5 & 2 & 2.5 & 3\\\hline
     Facebook & 1079 & 1063 & 1047 & 1037 & 1052 & 1047 \\
   Wiki & 1153	& 1166 & 1173 &	1213 &	1155 &	1167 \\
    HepPh & 967 & 1013 & 975 & 983 & 979 & 981 \\
   Enron & 2834 & 2764 & 2754 & 2777 & 2767 & 2762 \\
 \hline
	\end{tabular}}
    \vspace{-0.5cm}
\end{table}

\textbf{Running time}.
Finally, we evaluate \textbf{Q3} by testing the performance of CARGO and competitors while changing $n$.
Due to limited space, we only present the results of two graphs.
Fig. \ref{fig:time_facebook_n} and Fig. \ref{fig:time_wiki_n} show the execution time over Facebook and Wiki, respectively.
We can observe that the time cost of all methods grows with graph size $n$ increases, which is consistent with our theoretical analysis in Section \ref{sec:theoretical analysis}.
The running time of $\mathsf{Local2Rounds_\triangle}$ is approximately 2$\times$ higher than that of $\mathsf{CentralLap_\triangle}$. 
This is because $\mathsf{Local2Rounds_\triangle}$ needs an additional round for collecting a noisy global graph.
Another important observation is that CARGO needs more time overhead than the other two methods. 
For example, in Fig. \ref{fig:time_facebook_n}, CARGO takes the time of 485s while $\mathsf{Local2Rounds_\triangle}$ takes 0.235s and $\mathsf{CentralLap_\triangle}$ only takes 0.105s.
After further evaluation, we find that most of time overhead in CARGO is from the computation of secure triangle counting, namely, $\mathsf{Count}$ in Algorithm \ref{alg:local triangle counting}.
As shown in Fig. \ref{fig:time_wiki_n}, the execution time of $\mathsf{Count}$ accounts for at least 90\% of overall running time.
This is because CARGO needs to traverse all triples and the time complexity of triangle counting is up to $O(n^3)$.

\textbf{Summary of Experimental Results}.
In summary, our answers to three questions at the start of Section \ref{sec:experiment} are as follows.
\textbf{Q1}: Our CARGO outperforms significantly than $\mathsf{Local2Rounds_\triangle}$ and achieves comparative accuracy to $\mathsf{CentralLap_\triangle}$ without a trusted server.
\textbf{Q2}: Our $\mathsf{Project}$ algorithm significantly reduces the projection loss compared with the existing $\mathsf{GraphProjection}$.
\textbf{Q3}: The running time of CARGO is higher than that of other methods and most of computation overhead is from $\mathsf{Count}$.

\begin{figure*}[t]
	\centering
	\subfigure[Facebook]{
		\begin{minipage}[t]{0.238\linewidth}
			\centering
			\includegraphics[width=\linewidth]{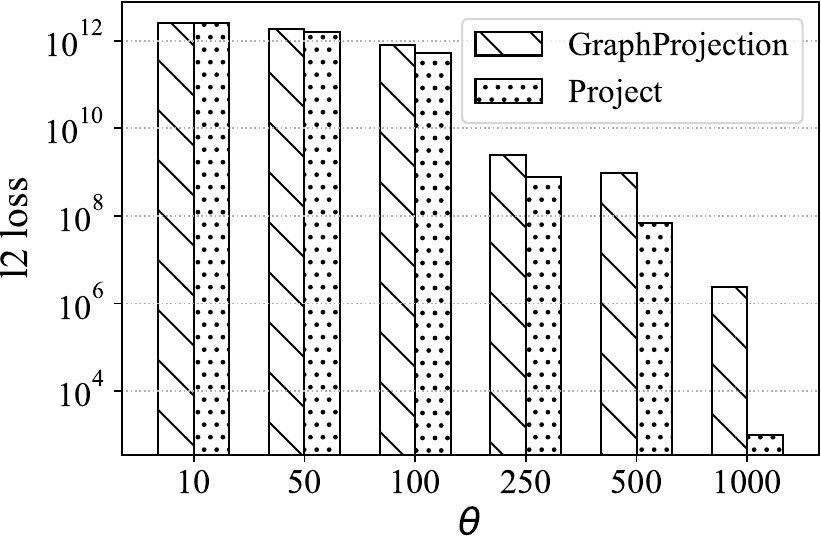}
            \label{fig:proloss_l2loss_facebook}
		\end{minipage}
	}%
	\subfigure[Wiki]{
		\begin{minipage}[t]{0.238\linewidth}
			\centering
			\includegraphics[width=\linewidth]{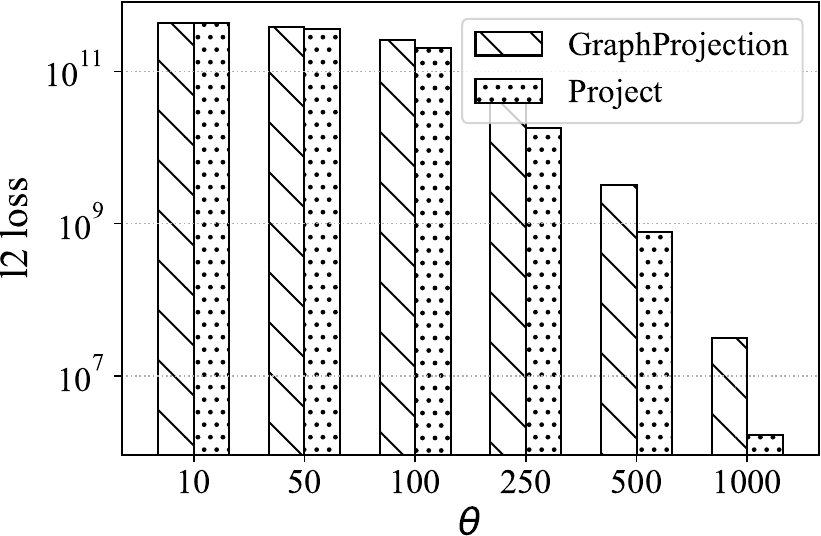}
		\end{minipage}%
	}
 	\subfigure[HepPh]{
		\begin{minipage}[t]{0.238\linewidth}
			\centering
			\includegraphics[width=\linewidth]{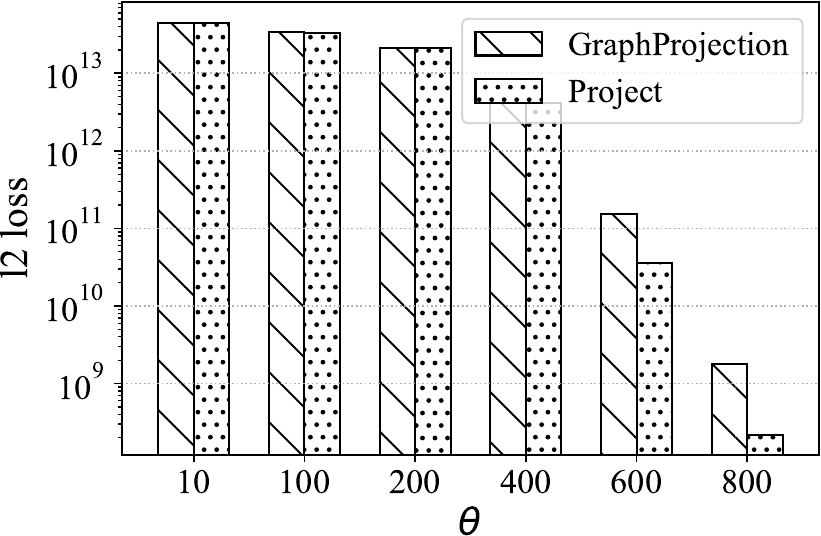}
		\end{minipage}%
	}
	\subfigure[Enron]{
		\begin{minipage}[t]{0.238\linewidth}
			\centering
			\includegraphics[width=\linewidth]{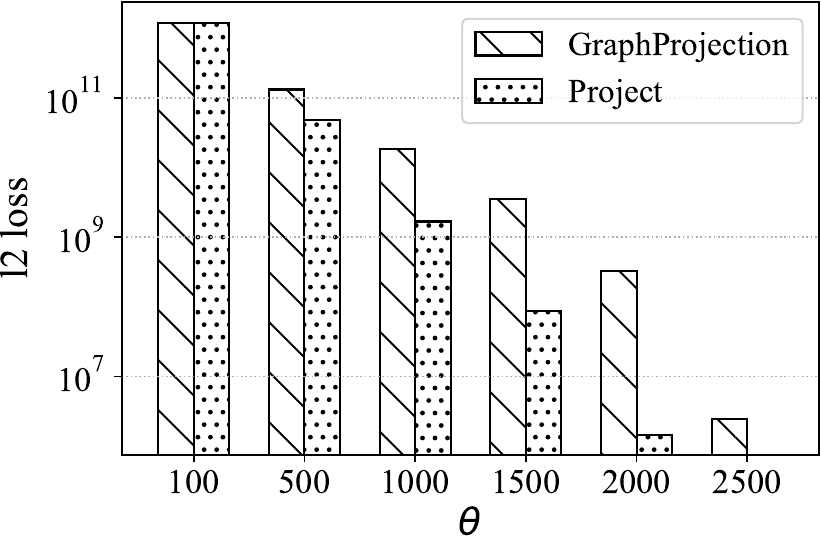}
		\end{minipage}%
	}
\centering
   \vspace{-0.5cm}
\caption{The $l_2$ loss of projection loss with  various parameters.}
   \vspace{-0.5cm}
\label{fig:proloss_l2loss}
\end{figure*}

\begin{figure*}[t]
	\centering
	\subfigure[Facebook]{
		\begin{minipage}[t]{0.238\linewidth}
			\centering
			\includegraphics[width=\linewidth]{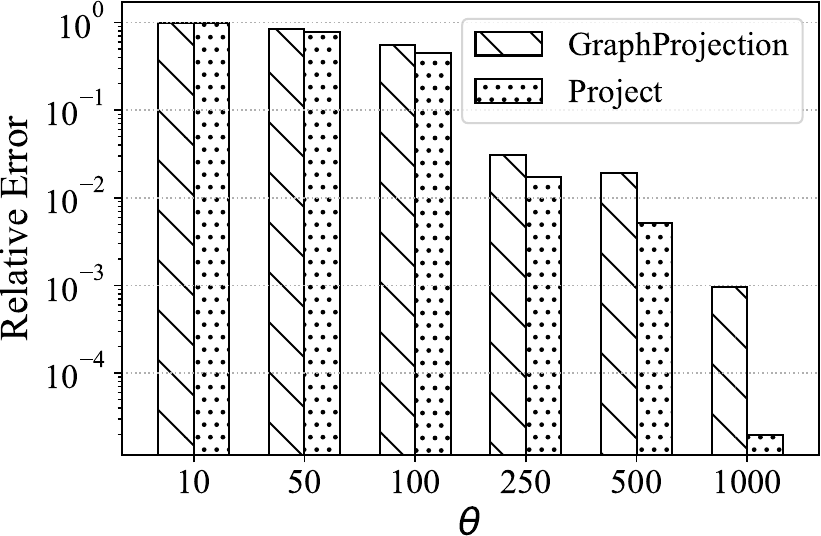}
		\end{minipage}
	}%
	\subfigure[Wiki]{
		\begin{minipage}[t]{0.238\linewidth}
			\centering
			\includegraphics[width=\linewidth]{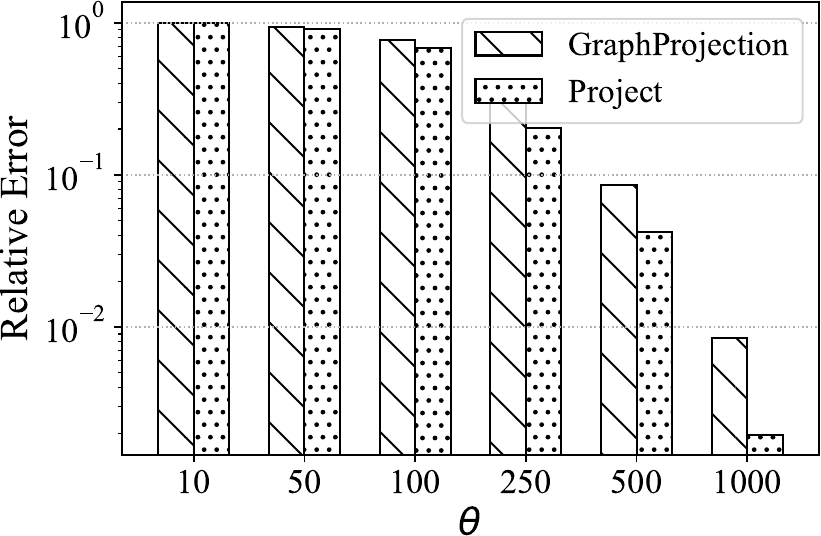}
		\end{minipage}%
	}
 	\subfigure[HepPh]{
		\begin{minipage}[t]{0.238\linewidth}
			\centering
			\includegraphics[width=\linewidth]{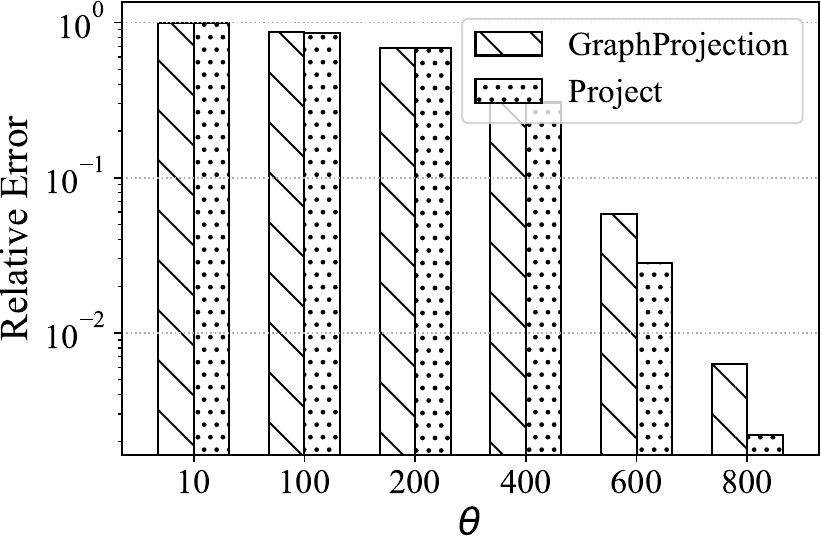}
		\end{minipage}%
	}
	\subfigure[Enron]{
		\begin{minipage}[t]{0.238\linewidth}
			\centering
			\includegraphics[width=\linewidth]{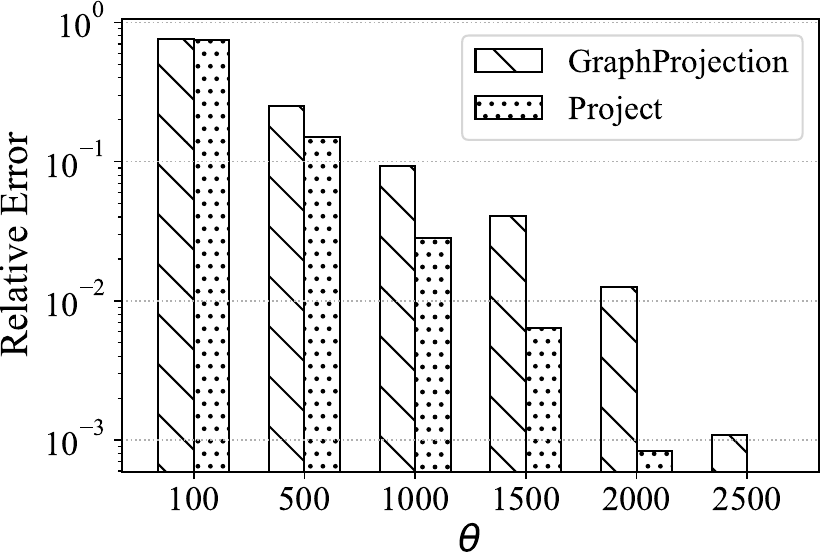}
		\end{minipage}%
	}
\centering
   \vspace{-0.5cm}
\caption{The relative error of projection loss with various parameters.}
   \vspace{-0.5cm}
\label{fig:proloss_re}
\end{figure*}

\begin{figure}
 \begin{center}
     \begin{minipage}[t]{0.48\linewidth}
   \centering
   \includegraphics[width=\linewidth]{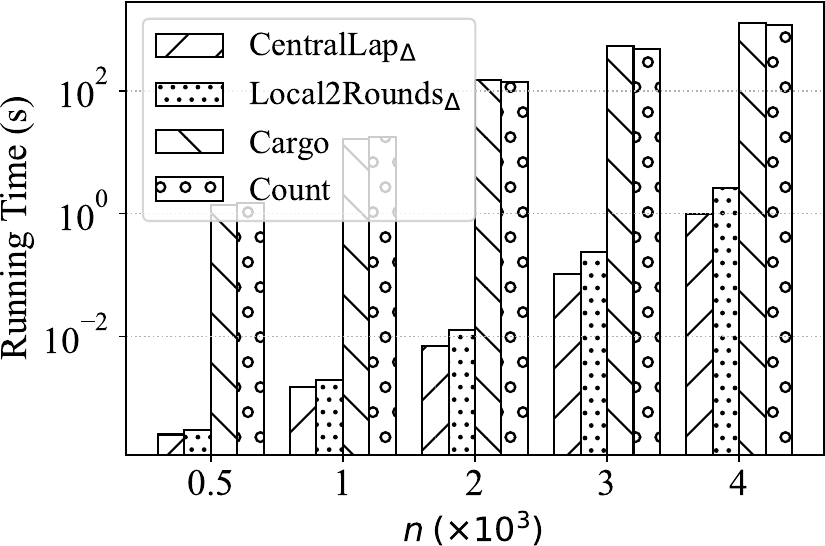}
    \vspace{-0.5cm}
   \caption{Running time on Facebook.}
    \vspace{-0.5cm}
   \label{fig:time_facebook_n}
\end{minipage}
  \begin{minipage}[t]{0.48\linewidth}
  \centering
   \includegraphics[width=\linewidth]{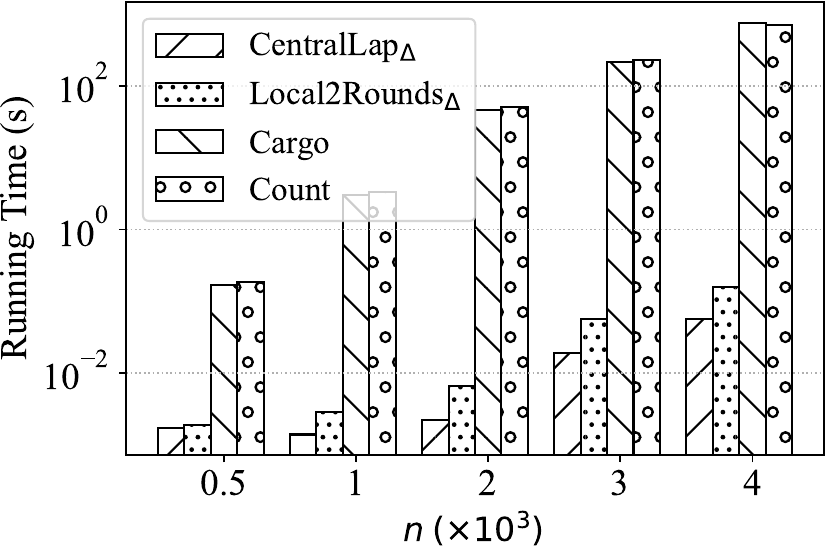}
    \vspace{-0.5cm}
  \caption{Running time on Wiki.}
  \label{fig:time_wiki_n}
  \end{minipage}
   \end{center}
 \vspace{-0.5cm}
\end{figure}

\section{Related Works}
\label{sec:related works}

\subsection{Triangle Counting in DP}
Differentially private triangle counting has been widely studied, and previous works are mainly based on either central DP (CDP) or local DP (LDP).

\textbf{Triangle Counting in CDP}.
Existing works related to triangle counting in the central model mainly focus on how to reduce the global sensitivity.
Ding et al. \cite{ding2021differentially} propose novel projection methods, namely, two edge-deletion strategies (DL and DS) for triangle counts distributions (histogram of triangle counts and cumulative histogram of triangle counts) while satisfying Node DP.
Karwa et al. \cite{karwa2011private} give a differentially private algorithm for releasing the triangle counts based on the higher-order local sensitivity, which adds less noise than methods with global sensitivity.
Kasiviswanathan et al. \cite{kasiviswanathan2013analyzing} develop algorithms for private graph analysis under Node DP.
They design a projection operator that projects the input graph into a bounded-degree (low-degree) graph, resulting in a lower sensitivity.
However, these works assume that there is a trusted server that owns the entire graph data. This, unfortunately, may not be practical in many applications due to the risk of privacy leaks \cite{yang2020local}.

\textbf{Triangle Counting in LDP}.
Triangle counting in LDP has recently attracted much attention. 
The main challenge is from the complex inter-dependencies involving multiple people. 
For example, each user cannot see edges between other users.
Imola et al. \cite{imola2021locally} propose a two-round interaction for triangle counts to handle the above challenge.
However, local randomization aggregates many errors and two-round interaction brings additional communication overhead.
Sun et al. \cite{sun2019analyzing} assume that each user has an extended view, and thus each user can see the third edges between others.
But this assumption may not hold in many graphs, and we make a minimal assumption where each user only knows her friends.
Also, locally added noise in this work influences the utility significantly.

In a nutshell, there is an apparent tradeoff between in CDP and in LDP in terms of privacy and utility while counting triangles.

\subsection{Crypto-assisted DP}
The approach of using cryptographic tools to enhance the utility of differential privacy (we call it as \textit{crypto-assistend DP}) has been studied in the literature\cite{he2017composing,gu2021precad,stevens2022efficient, truex2019hybrid,roy2020crypt,roth2019honeycrisp,fu2022dp,SunL21}.
He \textit{et al.} \cite{he2017composing} compose differential privacy and two-party computation for private record linkage while ensuring three
desiderata: correctness, privacy, and efficiency.
\colorR{Bohler et al.} proposed Crypt$\varepsilon$ \cite{roy2020crypt} is a system and a programming framework for supporting a rich class of  state-of-the-art DP programs, and it achieves the accuracy of CDP without a trusted server.
Honeycrisp \cite{roth2019honeycrisp} combines cryptographic techniques and differential privacy for answering periodic queries, which can sustainably run queries like the one from Apple’s deployment
while protecting user privacy in the long run.
Some works such as \cite{gu2021precad,stevens2022efficient,truex2019hybrid, SunL21} combine cryptography and differential privacy for federated learning, reducing the injected noise without sacrificing privacy.
Fu et al. \cite{fu2022dp} proposes a crypto-assisted differentially private framework for hierarchical count histograms under untrusted servers.
These protocol, therefore, achieve nearly the same accuracy as CDP models with untrusted servers.

Although the crypto-assisted DP model has been applied to tabular data \cite{he2017composing,roy2020crypt,roth2019honeycrisp,fu2022dp} and gradients in federated learning \cite{gu2021precad,stevens2022efficient,truex2019hybrid,SunL21},
characteristics of graph data, such as high-dimensional and
inter-correlated, lead to more challenges as discussed in
Section~\ref{sec:introduction}. To our knowledge, our work is the first work for triangle counting using crypto-assisted DP model.

\section{Conclusions}
\label{sec:conclusion}
To conclude, we propose the first crypto-assisted differentially private graph analysis framework, CARGO, which achieves high-utility triangle counting comparable to CDP-based models but without requiring a trusted server like LDP-based models.
Through theoretical and experimental analysis, we verify the privacy and utility achieved by our framework.

\section*{Acknowledgment}
This work was supported by JST SPRING (No. JPMJSP2110), JST CREST (No. JPMJCR21M2), JSPS KAKENHI (No. JP22H00521, JP22H03595, JP21K19767), and JST/NSF Joint Research SICORP (No. JPMJSC2107).
Jinfei Liu is the corresponding author.

\bibliographystyle{IEEEtran}
\bibliography{IEEEabrv,Reference}

\begin{thebibliography}{10}
\providecommand{\url}[1]{#1}
\csname url@samestyle\endcsname
\providecommand{\newblock}{\relax}
\providecommand{\bibinfo}[2]{#2}
\providecommand{\BIBentrySTDinterwordspacing}{\spaceskip=0pt\relax}
\providecommand{\BIBentryALTinterwordstretchfactor}{4}
\providecommand{\BIBentryALTinterwordspacing}{\spaceskip=\fontdimen2\font plus
\BIBentryALTinterwordstretchfactor\fontdimen3\font minus \fontdimen4\font\relax}
\providecommand{\BIBforeignlanguage}[2]{{%
\expandafter\ifx\csname l@#1\endcsname\relax
\typeout{** WARNING: IEEEtran.bst: No hyphenation pattern has been}%
\typeout{** loaded for the language `#1'. Using the pattern for}%
\typeout{** the default language instead.}%
\else
\language=\csname l@#1\endcsname
\fi
#2}}
\providecommand{\BIBdecl}{\relax}
\BIBdecl

\bibitem{seshadhri2019scalable}
C.~Seshadhri and S.~Tirthapura, ``Scalable subgraph counting: the methods behind the madness,'' in \emph{Companion Proceedings of The 2019 World Wide Web Conference}, 2019, pp. 1317--1318.

\bibitem{newman2009random}
M.~E. Newman, ``Random graphs with clustering,'' \emph{Physical review letters}, vol. 103, no.~5, p. 058701, 2009.

\bibitem{schank2005approximating}
T.~Schank and D.~Wagner, ``Approximating clustering coefficient and transitivity.'' \emph{Journal of Graph Algorithms and Applications}, vol.~9, no.~2, pp. 265--275, 2005.

\bibitem{goldsmith1990assessing}
T.~GOLDSMITH, ``Assessing structural similarity of graphs,'' \emph{Pathfinder Associative Networks: Studies in Knowledge Organization}, pp. 75--87, 1990.

\bibitem{tai2011privacy}
C.-H. Tai, P.~S. Yu, D.-N. Yang, and M.-S. Chen, ``Privacy-preserving social network publication against friendship attacks,'' in \emph{Proceedings of the 17th ACM SIGKDD international conference on Knowledge discovery and data mining}, 2011, pp. 1262--1270.

\bibitem{dwork2014algorithmic}
C.~Dwork, A.~Roth \emph{et~al.}, ``The algorithmic foundations of differential privacy,'' \emph{Foundations and Trends{\textregistered} in Theoretical Computer Science}, vol.~9, no. 3--4, pp. 211--407, 2014.

\bibitem{li2016differential}
N.~Li, M.~Lyu, D.~Su, and W.~Yang, ``Differential privacy: From theory to practice,'' \emph{Synthesis Lectures on Information Security, Privacy, \& Trust}, vol.~8, no.~4, pp. 1--138, 2016.

\bibitem{ding2021differentially}
X.~Ding, S.~Sheng, H.~Zhou, X.~Zhang, Z.~Bao, P.~Zhou, and H.~Jin, ``Differentially private triangle counting in large graphs,'' \emph{IEEE Transactions on Knowledge and Data Engineering}, vol.~34, no.~11, pp. 5278--5292, 2021.

\bibitem{karwa2011private}
V.~Karwa, S.~Raskhodnikova, A.~Smith, and G.~Yaroslavtsev, ``Private analysis of graph structure,'' \emph{Proceedings of the VLDB Endowment}, vol.~4, no.~11, pp. 1146--1157, 2011.

\bibitem{kasiviswanathan2013analyzing}
S.~P. Kasiviswanathan, K.~Nissim, S.~Raskhodnikova, and A.~Smith, ``Analyzing graphs with node differential privacy,'' in \emph{Theory of Cryptography: 10th Theory of Cryptography Conference, TCC 2013, Tokyo, Japan, March 3-6, 2013. Proceedings}.\hskip 1em plus 0.5em minus 0.4em\relax Springer, 2013, pp. 457--476.

\bibitem{imola2021locally}
J.~Imola, T.~Murakami, and K.~Chaudhuri, ``Locally differentially private analysis of graph statistics.'' in \emph{USENIX Security Symposium}, 2021, pp. 983--1000.

\bibitem{sun2019analyzing}
H.~Sun, X.~Xiao, I.~Khalil, Y.~Yang, Z.~Qin, H.~Wang, and T.~Yu, ``Analyzing subgraph statistics from extended local views with decentralized differential privacy,'' in \emph{Proceedings of the 2019 ACM SIGSAC Conference on Computer and Communications Security}, 2019, pp. 703--717.

\bibitem{imola2022communication}
J.~Imola, T.~Murakami, and K.~Chaudhuri, ``Communication-efficient triangle counting under local differential privacy,'' in \emph{31st USENIX Security Symposium (USENIX Security 22)}, 2022, pp. 537--554.

\bibitem{ye2020lf}
Q.~Ye, H.~Hu, M.~H. Au, X.~Meng, and X.~Xiao, ``Lf-gdpr: A framework for estimating graph metrics with local differential privacy,'' \emph{IEEE Transactions on Knowledge and Data Engineering}, vol.~34, no.~10, pp. 4905--4920, 2020.

\bibitem{he2017composing}
X.~He, A.~Machanavajjhala, C.~Flynn, and D.~Srivastava, ``Composing differential privacy and secure computation: A case study on scaling private record linkage,'' in \emph{Proceedings of the 2017 ACM SIGSAC Conference on Computer and Communications Security}, 2017, pp. 1389--1406.

\bibitem{roy2020crypt}
A.~Roy~Chowdhury, C.~Wang, X.~He, A.~Machanavajjhala, and S.~Jha, ``Crypt$\varepsilon$: Crypto-assisted differential privacy on untrusted servers,'' in \emph{Proceedings of the 2020 ACM SIGMOD International Conference on Management of Data}, 2020, pp. 603--619.

\bibitem{roth2019honeycrisp}
E.~Roth, D.~Noble, B.~H. Falk, and A.~Haeberlen, ``Honeycrisp: large-scale differentially private aggregation without a trusted core,'' in \emph{Proceedings of the 27th ACM Symposium on Operating Systems Principles}, 2019, pp. 196--210.

\bibitem{gu2021precad}
X.~Gu, M.~Li, and L.~Xiong, ``Precad: Privacy-preserving and robust federated learning via crypto-aided differential privacy,'' \emph{arXiv preprint arXiv:2110.11578}, 2021.

\bibitem{stevens2022efficient}
T.~Stevens, C.~Skalka, C.~Vincent, J.~Ring, S.~Clark, and J.~Near, ``Efficient differentially private secure aggregation for federated learning via hardness of learning with errors,'' in \emph{31st USENIX Security Symposium (USENIX Security 22)}, 2022, pp. 1379--1395.

\bibitem{truex2019hybrid}
S.~Truex, N.~Baracaldo, A.~Anwar, T.~Steinke, H.~Ludwig, R.~Zhang, and Y.~Zhou, ``A hybrid approach to privacy-preserving federated learning,'' in \emph{Proceedings of the 12th ACM workshop on artificial intelligence and security}, 2019, pp. 1--11.

\bibitem{SunL21}
L.~Sun and L.~Lyu, ``Federated model distillation with noise-free differential privacy,'' in \emph{Proceedings of the Thirtieth International Joint Conference on Artificial Intelligence, {IJCAI}}, 2021, pp. 1563--1570.

\bibitem{fu2022dp}
C.~Fu, H.~Li, J.~Lou, and J.~Cui, ``Dp-horus: Differentially private hierarchical count histograms under untrusted server,'' in \emph{Proceedings of the 31st ACM International Conference on Information \& Knowledge Management}, 2022, pp. 530--539.

\bibitem{day2016publishing}
W.-Y. Day, N.~Li, and M.~Lyu, ``Publishing graph degree distribution with node differential privacy,'' in \emph{Proceedings of the 2016 International Conference on Management of Data}, 2016, pp. 123--138.

\bibitem{durak2012degree}
N.~Durak, A.~Pinar, T.~G. Kolda, and C.~Seshadhri, ``Degree relations of triangles in real-world networks and graph models,'' in \emph{Proceedings of the 21st ACM international conference on Information and knowledge management}, 2012, pp. 1712--1716.

\bibitem{shamir1979share}
A.~Shamir, ``How to share a secret,'' \emph{Communications of the ACM}, vol.~22, no.~11, pp. 612--613, 1979.

\bibitem{shi2011privacy}
E.~Shi, H.~Chan, E.~Rieffel, R.~Chow, and D.~Song, ``Privacy-preserving aggregation of time-series data,'' in \emph{Annual Network \& Distributed System Security Symposium (NDSS)}.\hskip 1em plus 0.5em minus 0.4em\relax Internet Society., 2011.

\bibitem{acs2011have}
G.~{\'A}cs and C.~Castelluccia, ``I have a dream!(differentially private smart metering).'' in \emph{Information hiding}, vol. 6958.\hskip 1em plus 0.5em minus 0.4em\relax Springer, 2011, pp. 118--132.

\bibitem{goryczka2015comprehensive}
S.~Goryczka and L.~Xiong, ``A comprehensive comparison of multiparty secure additions with differential privacy,'' \emph{IEEE transactions on dependable and secure computing}, vol.~14, no.~5, pp. 463--477, 2015.

\bibitem{patra2021aby2}
A.~Patra, T.~Schneider, A.~Suresh, and H.~Yalame, ``Aby2. 0: Improved mixed-protocol secure two-party computation.'' in \emph{USENIX Security Symposium}, 2021, pp. 2165--2182.

\bibitem{rathee2020cryptflow2}
D.~Rathee, M.~Rathee, N.~Kumar, N.~Chandran, D.~Gupta, A.~Rastogi, and R.~Sharma, ``Cryptflow2: Practical 2-party secure inference,'' in \emph{Proceedings of the 2020 ACM SIGSAC Conference on Computer and Communications Security}, 2020, pp. 325--342.

\bibitem{murakami2019utility}
T.~Murakami and Y.~Kawamoto, ``Utility-optimized local differential privacy mechanisms for distribution estimation,'' in \emph{28th USENIX Security Symposium (USENIX Security 19)}, 2019, pp. 1877--1894.

\bibitem{wang2017locally}
T.~Wang, J.~Blocki, N.~Li, and S.~Jha, ``Locally differentially private protocols for frequency estimation,'' in \emph{26th USENIX Security Symposium (USENIX Security 17)}, 2017, pp. 729--745.

\bibitem{chen2012differentially}
R.~Chen, G.~Acs, and C.~Castelluccia, ``Differentially private sequential data publication via variable-length n-grams,'' in \emph{Proceedings of the 2012 ACM conference on Computer and communications security}, 2012, pp. 638--649.

\bibitem{bindschaedler2016synthesizing}
V.~Bindschaedler and R.~Shokri, ``Synthesizing plausible privacy-preserving location traces,'' in \emph{2016 IEEE Symposium on Security and Privacy (SP)}.\hskip 1em plus 0.5em minus 0.4em\relax IEEE, 2016, pp. 546--563.

\bibitem{hay2009accurate}
M.~Hay, C.~Li, G.~Miklau, and D.~Jensen, ``Accurate estimation of the degree distribution of private networks,'' in \emph{2009 Ninth IEEE International Conference on Data Mining}.\hskip 1em plus 0.5em minus 0.4em\relax IEEE, 2009, pp. 169--178.

\bibitem{raskhodnikova2016differentially}
S.~Raskhodnikova and A.~Smith, ``Differentially private analysis of graphs,'' \emph{Encyclopedia of Algorithms}, 2016.

\bibitem{qin2017generating}
Z.~Qin, T.~Yu, Y.~Yang, I.~Khalil, X.~Xiao, and K.~Ren, ``Generating synthetic decentralized social graphs with local differential privacy,'' in \emph{Proceedings of the 2017 ACM SIGSAC Conference on Computer and Communications Security}, 2017, pp. 425--438.

\bibitem{mohassel2017secureml}
P.~Mohassel and Y.~Zhang, ``Secureml: A system for scalable privacy-preserving machine learning,'' in \emph{2017 IEEE symposium on security and privacy (SP)}.\hskip 1em plus 0.5em minus 0.4em\relax IEEE, 2017, pp. 19--38.

\bibitem{riazi2018chameleon}
M.~S. Riazi, C.~Weinert, O.~Tkachenko, E.~M. Songhori, T.~Schneider, and F.~Koushanfar, ``Chameleon: A hybrid secure computation framework for machine learning applications,'' in \emph{Proceedings of the 2018 on Asia conference on computer and communications security}, 2018, pp. 707--721.

\bibitem{zheng2023secure}
S.~Zheng, Y.~Cao, and M.~Yoshikawa, ``Secure shapley value for cross-silo federated learning,'' \emph{Proceedings of the VLDB Endowment}, vol.~16, no.~7, pp. 1657--1670, 2023.

\bibitem{liucrypto}
S.~Liu, Y.~Cao, T.~Murakami, and M.~Yoshikawa, ``A crypto-assisted approach for publishing graph statistics with node local differential privacy,'' in \emph{2022 IEEE International Conference on Big Data (Big Data)}, 2022, pp. 5765--5774.

\bibitem{rabin2005exchange}
M.~O. Rabin, ``How to exchange secrets with oblivious transfer,'' \emph{Cryptology ePrint Archive}, 2005.

\bibitem{kilian1988founding}
J.~Kilian, ``Founding crytpography on oblivious transfer,'' in \emph{Proceedings of the twentieth annual ACM symposium on Theory of computing}, 1988, pp. 20--31.

\bibitem{kotz2001laplace}
S.~Kotz, T.~Kozubowski, and K.~Podg{\'o}rski, \emph{The Laplace distribution and generalizations: a revisit with applications to communications, economics, engineering, and finance}.\hskip 1em plus 0.5em minus 0.4em\relax Springer Science \& Business Media, 2001, no. 183.

\bibitem{lindell2017simulate}
Y.~Lindell, ``How to simulate it--a tutorial on the simulation proof technique,'' \emph{Tutorials on the Foundations of Cryptography: Dedicated to Oded Goldreich}, pp. 277--346, 2017.

\bibitem{murphy2012machine}
K.~P. Murphy, \emph{Machine learning: a probabilistic perspective}.\hskip 1em plus 0.5em minus 0.4em\relax MIT press, 2012.

\bibitem{dong2022nearly}
W.~Dong and K.~Yi, ``A nearly instance-optimal differentially private mechanism for conjunctive queries,'' in \emph{Proceedings of the 41st ACM SIGMOD-SIGACT-SIGAI Symposium on Principles of Database Systems}, 2022, pp. 213--225.

\bibitem{dong2021residual}
{W. Dong and K. Yi}, ``Residual sensitivity for differentially private multi-way joins,'' in \emph{SIGMOD/PODS'21: Proceedings of the 2021 International Conference on Management of Data}, 2021.

\bibitem{snapnets}
J.~Leskovec and A.~Krevl, ``{SNAP Datasets}: {Stanford} large network dataset collection,'' \url{http://snap.stanford.edu/data}, Jun. 2014.

\bibitem{yang2020local}
M.~Yang, L.~Lyu, J.~Zhao, T.~Zhu, and K.-Y. Lam, ``Local differential privacy and its applications: A comprehensive survey,'' \emph{arXiv preprint arXiv:2008.03686}, 2020.

\end{thebibliography}

\end{document}